\documentclass[11pt]{article}

\usepackage{url}
\usepackage[bottom]{footmisc}
\usepackage[caption=false,font=footnotesize]{subfig}
\usepackage{amsmath}
\usepackage{amssymb,hyperref,graphicx}
\usepackage{amsfonts,multirow}
\usepackage[left=1.5cm, right=2cm]{geometry}

\begin{document}

\title{On Networks and their Applications: Stability of Gene Regulatory Networks and Gene Function Prediction using Autoencoders} 
\author{Hamza Çoban} 

\maketitle

\begin{abstract}
    We prove that nested canalizing functions are the
  minimum-sensitivity Boolean functions for any activity ratio and we determine the functional form of this boundary which has a nontrivial fractal structure. We further observe that the majority of the gene regulatory functions found in known biological networks (submitted to the Cell Collective database) lie on the line of minimum sensitivity which paradoxically remains largely in the unstable regime. Our results provide a quantitative basis for the argument that an evolutionary preference for nested canalizing functions in gene regulation (e.g., for higher robustness) and for elasticity of gene activity are sufficient for concentration of such systems near the ``edge of chaos.'' The original structure of gene regulatory networks is unknown due to the undiscovered functions of some genes. Most gene function discovery approaches make use of unsupervised clustering or classification methods that discover and exploit patterns in gene expression profiles. However, existing knowledge in the field derives from multiple and diverse sources. Incorporating this know-how for novel gene function prediction can, therefore, be expected to improve such predictions. We here propose a function-specific novel gene discovery tool that uses a semi-supervised autoencoder. Our method is thus able to address the needs of a modern researcher whose expertise is typically confined to a specific functional domain. Lastly, the dynamics of unorthodox learning approaches like biologically plausible learning algorithms are investigated and found to exhibit a general form of Einstein relation.
\end{abstract}

\tableofcontents
\listoftables
\listoffigures

\part{Dynamics Stability Analysis of Gene Regulatory Networks}
\label{chapter:format}
\section{Introduction and Literature Review}

Genes are the determinants of cell fate. They encode and regulate protein production required for almost all biological processes. The type of a cell (e.g., a neuron or a muscle cell) and its characteristics (e.g., healthy or cancerous) are determined by the activity profile of the organism's genes in that particular cell. While some biological processes require an environmental signal that triggers the activation of a gene, some others involve genes that are turned on and off repeatedly in a cycle. Most of the time, several genes work together in order to fulfill a particular task. Consequently, a communication network facilitating cooperative action of  genes is inevitable. Gene regulatory networks constitute the infrastructure on which the coupling between gene expression and gene-gene interactions is exploited by Nature in order to maintain the cell's proper functioning and responsiveness to environmental changes. 

Gene regulatory networks (GRNs) have been of interest for some time in disciplines other than biology, such as, physics, mathematics and computer science. Biological studies elucidated the mechanisms of gene regulation and its inherent network structure, through which the developmental differentiation of identical cells and how cells maintain biological processes can be understood. In the reviews \cite{Davidson2008,Levine2005}, the role and properties of GRNs in the formation of body parts is described for several organisms. Another review \cite{erwin2009evolution} investigates the evolution of the hierarchical structure of GRNs and how they shape the morphology of the associated organisms.

On the computational/theoretical side, methods for modelling GRNs have been discussed recently in \cite{Huynh-Thu2019, Chai2014}. Two conventional approaches for simulating GRNs are (graph-based) network models and systems of differential equations. The main difference between the two is that the networks are typically built on discrete-time Markovian dynamics with a Boolean state space, while the differential equation models are continuous both in time and state representation. The dynamic state of a gene regulation network can be either probabilistic or deterministic, depending on the modeling approach. \cite{Chai2014} discuss the advantages and weaknesses of different approaches comparatively. More recently, controllability of gene regulation has been of much interest as reflected in \cite{borriello2021basis,gates2016control,park2019dynamic,toyoda2019optimal}.

Numerous studies have addressed the dynamics and stability of Boolean networks. First seminal papers on the stability of GRNs are by S.A. Kauffman: In his pioneering work \cite{kauffman1969metabolic}, he discovered that randomly constructed Boolean network models of GRNs with fixed connectivity K=2 exhibit stable patterns. He further argued that genetic nets have an inherent order due to having low levels of connectivity, and they remain in an ordered (i.e., dynamically robust) state while, at the same time, being able to respond to the environmental changes. The stability of GRNs has been observed in several organisms later with the development of microarray technology \cite{kauffman2003random,kauffman2004genetic,shmulevich2005eukaryotic}.

In subsequent studies, the focus of the discussion on stability shifted towards criticality (in the sense used in statistical physics and phase transitions). It was suggested that the biological systems from the molecular level to the organism level reside in the critical boundary situated between stable and unstable regimes \cite{mora2011biological}. The critical phenomena exhibited by GRNs were observed and theoretically discussed in, for example, \cite{fourkingdoms,shmulevich2005eukaryotic}. Critical Boolean networks emerge from basic evolution models, as shown in multiple studies which strengthen the claim of self-organized criticality  set through natural evolution in GRNS \cite{bornholdt2000topological,liu2006emergent,ramo2006perturbation,torres2012criticality}. As an order parameter for Boolean networks, Shmulevich and Kauffman introduced the notion of "sensitivity" \cite{shmulevich2004activities}. Recently, a survey on the sensitivities of over 70 documented Boolean network models of GRNs demonstrated their criticality, a feature not shared by randomly generated graphs with similar global parameters \cite{daniels2018criticality}.

An important concept in studies of gene regulation is canalization. Canalization refers to an induced ability to acquire certain phenotypes or maintain the status of the cell regardless of environmental changes except the triggering event \cite{waddington1942canalization}. Since, by definition, it refers to a stability condition, few quantitative and rigorous studies that investigate the relation between canalization and stability have been published. Earlier works show that the expected stability of canalizing Boolean functions is higher than their random counterparts and it is possible to quantify the expected sensitivity of a canalizing function \cite{kauffman2003random,kauffman2004genetic}. Latter studies discuss the critical regimes of networks with canalizing functions. \cite{peixoto2010phase,moreira2005canalizing}. The special subclass of canalizing functions, nested canalizing functions (NCFs), were also investigated in the context of the sensitivity of GRNs. It was found that the expected sensitivity drops with increasing canalizing depth, although the dependence becomes rather weak beyond a certain depth \cite{layne2012nested}. Studies in the opposite direction also established an upper bound of 4/3 on the sensitivities of NCFs (with 1 corresponding to the critical order-chaos boundary) \cite{li2013boolean,klotz2013bounds}. A quantification of the expected sensitivities as a function of activities of partially nested canalizing functions was also done recently \cite{kadelka2017influence}. 

This chapter describes the findings reported in two recent articles by the author \cite{ccoban2022proof,ccoban2022critical} and additional discussions on gene regulatory networks.

\section{Gene Regulatory Networks}

The genetic material, DNA, is the machine code of a cell. To run the machine, proteins and their complexes are essential and some regions of DNA, genes, contain the recipe for specific proteins. The cell has regulatory mechanisms to control the protein levels since some proteins must be present inside the cytoplasm in a certain amount, some are occasional ones, like signal proteins, division-related proteins, and so on. Though cells of a single organism contain the same DNA, they specialize and differentiate to perform distinct tasks like  neural transmission, enzyme production, and killing bacterial intruders. Thus, different sets of proteins should be produced in each cell and the protein levels and distinct proteins actually define the state of a cell. This state can be regulated again by DNA itself, in this regard DNA constitutes the cell's brain, which responds to the current state of the cell and the environment through inducing necessary protein production by producing associated proteins. In this manner, this interaction between protein production is a network structure that may include cyclic patterns. 

In this genetic network, genes, mRNAs, proteins, and transcription factors (TFs) are involved. Genes are transcripted into mRNAs which is read by ribosomes for protein production. The rate of transcription of each gene is controlled via its promoter region. Transcription factors may blockade this region to prevent transcription or induce transcription by activating this region. TFs are again some proteins produced through the reading of mRNAs by ribosomes and some may include other complexes acquired from the environment. In short, genes are responsible for protein production and some proteins are responsible to regulate the transcription of genes. After an abstraction of protein interaction, the activation level of a gene is related to other genes' activation levels. 

Gene-gene interactions of this kind can be modeled with an ODE system, where each function represents the activity level of a gene. This method shows adequate results; however, it is hard to infer parameters with precision due to the lack of experiments in a continuous time scale and the cost of such experiments. The parameter space is also increasing with the network size and this is another downside of such models. Discrete network models are more appropriate for existing experiments on gene expression profiles. Among network models, Boolean networks promise decent prediction power and simplicity.

\section{GRNs as Boolean Networks and their Dynamics}

In the Boolean network model of GRNs, each node represents a gene. Each node has a state which depicts its activity and this state is restricted to a binary number, 0 or 1. Edges correspond to interaction between genes and there are two types of edges: inhibitory and promoting. The edges are directed since these interactions are not reciprocal by default.

\begin{figure}
    \centering
    \includegraphics{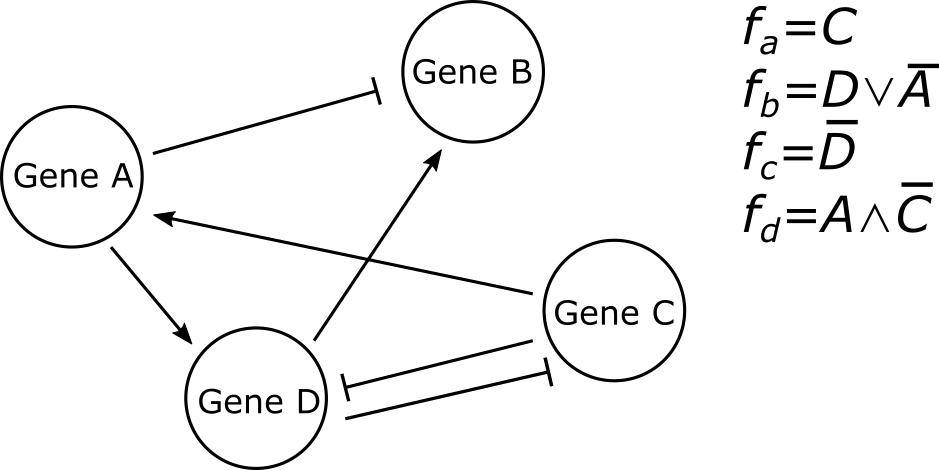}
    \caption{An example of a Boolean network with 4 genes and their associated Boolean functions. Edges with arrow represent promoting effect and the ones with the cutting end represent the inhibitory effect.}
    \label{fig:example_GRN}
\end{figure}

A Markovian dynamic is defined over the state space of a Boolean network. At each time step, the state of a node is determined with a Boolean function whose inputs are the incoming edges. In other words, each node, or gene, has an associated Boolean function that gives the activity of it at any time. In the synchronous update scheme, The activity of $i^{th}$ node at time $t+1$ is
$$x_i(t+1)=f_i(x_{\{I\}}(t)) \ ,$$
where $x$ is the binary state vector of the network and $I$ is the set of nodes which is the origin of a directed edge toward node $i$.

\section{Symmetry and Generalized Symmetry}

The symmetry exists in a Boolean function when two inputs of it are interchangeable in every possible state. In the mathematical language a Boolean function $f:\{0,1\}^n \xrightarrow{} \{0,1\}$ is symmetric if and only if there is an input pair $(x_i,x_j)$ such that $f(S_{ij}\Vec{x})=f(\Vec{x}), \ \forall \Vec{x}\in \{0,1\}^n$, where $S_{ij}$ is the exchange operator which exchanges the values of $x_i$ and $x_j$. 

 A Boolean function $f:\{0,1\}^n \xrightarrow{} \{0,1\}$ is generally symmetric if and only if there is an input pair $(x_i,x_j)$ such that $f(S_{ij}\Vec{x})=f(\Vec{x}), \ \forall \Vec{x}\in \{0,1\}^n$ or $f(S_{ij}N_i\Vec{x})=f(N_i\Vec{x}), \ \forall \Vec{x}\in \{0,1\}^n$, where $N_{i}$ is the negation operator which switches the value of $x_i$. 
 
 Consider two basic functions $f_1=x_1\vee x_2$ and $f_2=x_1\vee \Bar{x}_2$. With these definitions $f_1$ is symmetric but $f_2$ is not. However, their structure is identical and they are both generally symmetric. For sensitivity analysis, a more appropriate definition is generalized symmetry since in these two cases the sensitivity is the same. After this discussion symmetry refers to generalized symmetry throughout the paper.

 \subsection{2-symmetric functions}

 Symmetry is an equivalence relation on the set of inputs. Thus, equivalence classes exist within this context. A Boolean function is $2-$symmetric if and only if their inputs constitute two equivalence classes on symmetry. 

 Let $f$ be a Boolean function with $f:\{0,1\}^n \xrightarrow{} \{0,1\}$. If there is a partition on the input set, $(\{ x_1,x_2,\dots,x_a\},\{x_{a+1},\dots,x_n\})$ such that $f(S_{ij}\Vec{x})=f(\Vec{x})$ or $f(S_{ij}N_i\Vec{x})=f(N_i\Vec{x}), \ (\forall i,j \in \{1,2,\dots,a\} \text{ or } \forall i,j \in \{a+1,a+2,\dots,n\}),\ \forall \Vec{x}\in \{0,1\}^n$ .
 
 \noindent If we ignore the xor operator the canonical form of $2-$symmetric functions are
 \begin{equation}
f=g_1(x_1, \dots ,x_a) \vee g_2(x_{a+1}, \dots ,x_n) \ ,
\end{equation}
where $g_1$ and $g_2$ are totally symmetric functions. Such functions can be classified into four types according to their structure. 

\begin{eqnarray}
    Type-0 &\quad\mbox{, if } f=x_1 \vee \dots \vee x_n\nonumber\\
    Type-1&
    \quad\mbox{, if } f= (x_1 \wedge \dots \wedge x_a) \vee x_{a+1} \vee \dots \vee x_n \nonumber\\
    Type-2&
    \quad\mbox{, if } f= x_1 \vee \dots \vee x_a \vee (x_{a+1} \wedge \dots \wedge x_n)\nonumber \\
    Type-3&
    \quad\mbox{, if } f= (x_1 \wedge \dots \wedge x_a) \vee (x_{a+1} \wedge \dots \wedge x_n) \ .
    \label{eq:eq-types}
\end{eqnarray}
In this classification, the negations of each input and the total negation of a Boolean function are ignored since they are irrelevant for sensitivity analysis. For example,
\begin{align*}
    f&=x_1\wedge \Bar{x}_2 \wedge \Bar{x}_3 \ \xrightarrow{} \text{Type 0}\\
    f&=(x_1\vee \Bar{x}_2) \wedge \Bar{x}_3 \wedge \Bar{x}_4 \ \xrightarrow{} \text{Type 1 or Type 2}\\
    f&=(x_1\vee \Bar{x}_2) \wedge (x_3\vee \Bar{x}_4) \ \xrightarrow{} \text{Type 3}\\
\end{align*}

Another important notice is that type 1 and type 2 functions have essentially the same structure with different sizes of equivalence classes. For simplicity, the calculations are done separately for both of them.
\section{Canalization in Boolean Setting}

\subsection{Canalizing Functions}
In the Boolean context, canalization refers to the presence of an input or inputs which dictate the output of a function when they acquire their canalizing value regardless of the state of other inputs. Formally $f$ is canalizing in $x_1$ if
 \begin{equation}
  f(x_1,x_2, \dots, x_n) \equiv
  \begin{cases}
    r_1,\ \mbox{if}\ x_1 = \sigma_1 \\
    \hat{f}(x_{2},x_{3},\dots,x_n), \mbox{otherwise}\ ,
  \end{cases}
  \label{ncf_def}
\end{equation}
where $\hat{f}$ is an arbitrary Boolean function with $n-1$ input, $r_1$ is the canalized value or state and $\sigma_1$ is the canalizing value or state. 

It is possible to write the explicit version of canalizing functions. If canalized and canalizing states are 1, then the function $f$ has the following form, 
$$f(x_1,x_2,\dots,x_n)=x_1 \vee \hat{f}(x_2,x_3,\dots,x_n)$$
With the same strategy, one can write all possible forms of a canalizing function depending on its canalized and canalizing value.
\begin{equation}
    f(x_1,x_2,\dots,x_n)=
    \begin{cases}
        x_1 \vee \hat{f}(x_2,x_3,\dots,x_n), \ \mbox{if} (r_1,\sigma_1)=(1,1)\\
        \bar{x_1} \vee \hat{f}(x_2,x_3,\dots,x_n), \ \mbox{if} (r_1,\sigma_1)=(1,0)\\
        \bar{x_1} \wedge \hat{f}(x_2,x_3,\dots,x_n), \ \mbox{if} (r_1,\sigma_1)=(0,1)\\
        x_1 \wedge \hat{f}(x_2,x_3,\dots,x_n), \ \mbox{if} (r_1,\sigma_1)=(0,0)\\
    \end{cases}
    \label{explicit_cf}
\end{equation}

The activity $\alpha[f]$ is the number of input states which sets the function $f$ to 1. The activity ratio of $p[f]$ is then the fraction of these states among all possible states. With the definition above, if $r_1$ is 1, then half of the input states $(2^{n-1}-many)$ sets to output to 1. The output of the other half is given with the function $\hat{f}$. Then the activity ratio of a canalizing function is equal to $\alpha[f]=r_1\times 2^{n-1}+\alpha[\hat{f}]$. 

\subsection{Partially Nested Canalizing Functions}
In general, canalization may occur in several stages. In hierarchical order, the first input has the canalization power; but if it fails to have canalizing state, the power is acquired by the second input in the hierarchy. The rigorous definition for a partially nested canalizing function (pNCF) is
\begin{equation}
  f(x_1, \dots, x_n) \equiv
  \begin{cases}
    r_1,\ \mbox{if}\ x_1 = \sigma_1 \\
    r_2,\ \mbox{if}\ x_2 = \sigma_2, x_1=\bar{\sigma_1}\\
    \vdots \\
    r_k,\ \mbox{if}\ x_n = \sigma_n 
    \mbox{ and}\,x_i=\bar{\sigma_i}\ \forall i \in \{1,\dots ,n-1\}  \\
    \hat{f}(x_{k+1},\dots,x_n),\ \mbox{otherwise}\ ,
  \end{cases}
  \label{pncf_def}
\end{equation}
where $k$ is defined as canalizing depth. Using the pattern in Eq.\ref{explicit_cf}, the closed-form expression of a pNCF can be written. The method to write it consists of $k$ steps, and at each step, the form is revealed by the pair $(r_k,\sigma_k)$. For example if $r_1$ and $\sigma_1$ are 1, then $f=x_1 \vee f(x_1=0,x_2,\dots,x_n)$. Then, the reduced function can be further reduced by utilizing $r_2$ and $\sigma_2$.
The general closed-form expression of a pNCF is,
\begin{equation}
    f(x_1,x_2,\dots,x_n)=x_1^{(r_1,\sigma_1)} \ \diamondsuit_{r_1} ( x_2^{(r_2,\sigma_2)}\ \diamondsuit_{r_2} ( \dots ( x_k^{(r_k,\sigma_k)} \ \diamondsuit_{r_k} \hat{f}(x_{k+1},\dots,x_n))\dots)) \ ,
\end{equation}
where $x_i^{(0,0)}=x_i$, $x_i^{(0,1)}=\bar{x_i}$, $x_i^{(1,0)}=\bar{x_i}$, $x_i^{(1,1)}=x_i$, $\diamondsuit_0=\wedge$, and $\diamondsuit_1=\vee$

The activity of a pNCF can also be written in terms of canalized values and the function $\hat{f}$. As in the previous case, the first canalizing input determines the output of half of the input states. Then the second one determines the half of the remaining half $(2^{n-2}-many)$. Explicitly the activity of the pNCF $f$ is,
\begin{equation}
    \alpha[f]=\alpha[\hat{f}]+\sum\limits_{i=1}^k r_i\times 2^{n-k}
    \label{pncf_activity}
\end{equation}

\subsection{Nested Canalizing Functions}
An extreme case of canalization is when all inputs exert canalization in a (non-unique) hierarchical order:
\begin{equation}
  f(x_1, \dots, x_n) \equiv
  \begin{cases}
    r_1,\ \mbox{if}\ x_1 = \sigma_1 \\
    r_2,\ \mbox{if}\ x_2 = \sigma_2, x_1=\bar{\sigma_1}\\
    \vdots \\
    r_n,\ \mbox{if}\ x_n = \sigma_n 
    \mbox{ and}\,x_i=\bar{\sigma_i}\ \forall i \in \{1,\dots ,n-1\}  \\
    \bar{r}_n,\ \mbox{otherwise}\ .
  \end{cases}
  \label{ncf_def}
\end{equation}
In this equation, $r_i$ s are the canalized states, and $\sigma_i$ s are the canalizing values. Given a nested structure, an NCF can be reduced and obtain another NCF with one less input. If we set $x_1=\Bar{\sigma_1}$, $f(x_1=\Bar{\sigma_1},x_2,\dots,x_n)$ is also a NCF. Lastly, the activity of an NCF only depends on the canalized states $r_i$ s.
\begin{equation}
    \alpha[f]=\sum\limits_{i=1}^{n-1} r_i\times 2^{n-i}+1=(r_1r_2\dots r_{n-1}1)_2
    \label{ncf_p}
\end{equation}
Notice that the activity is always odd if there is no redundant input. The last two conditions of Eq.\ref{ncf_def} ensure that all inputs are relevant to decide the output of the function. If an NCF has an even activity, it definitely has at least one redundant input, and it is possible to remove them to obtain an NCF whose all inputs are relevant.

\begin{figure}[h!]
    \centering
    \includegraphics[width=0.9\textwidth]{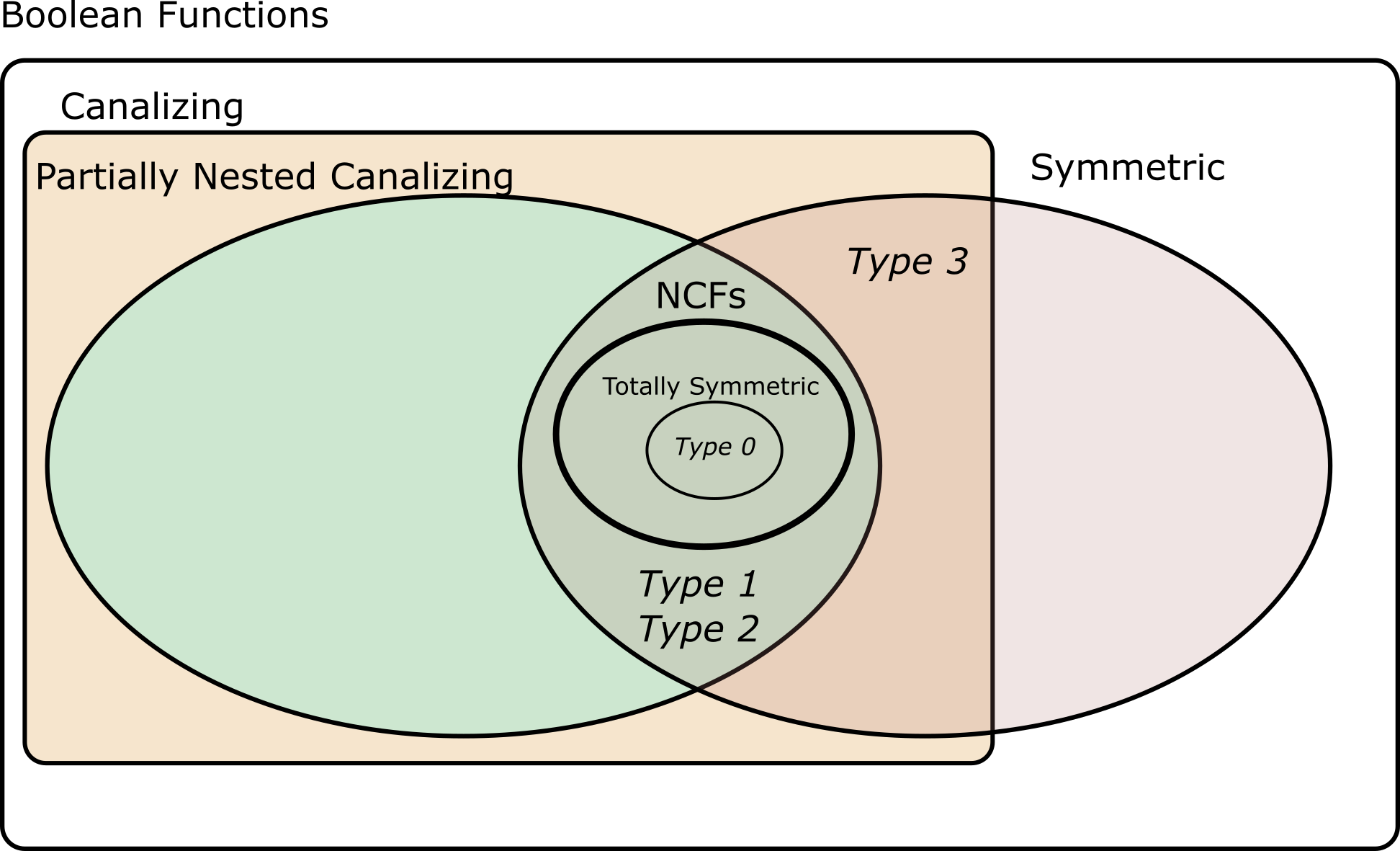}
    \caption{The Venn diagram depicting the relationship between various types of canalizing and symmetric functions}
    \label{fig:venn}
\end{figure}

A function can acquire both symmetry and canalization properties. Considering intersections of such cases, type-0 functions are always nested canalizing. Also, all nested canalizing functions have some symmetric pairs. The hierarchy of NCFs consist of layers, whose members have identical privilege and are interchangeable on the hierarchy list. For instance, considering the function $f=x_1 \vee (x_2 \wedge x_3)$, $x_1$ is on the top the hierarchy while $x_2$ and $x_3$ share the next layer. A nested canalizing function with $r$ layers is \textit{r-symmetric}. The connection between canalization and symmetry in the context of \textit{2-symmetric} functions is shown as a Venn diagram in Figure \ref{fig:venn}.

\section{Sensitivity in Boolean Networks}

For stability purposes, one can investigate Lyapunov exponents of GRNs. Lyapunov exponent gives a measure of how much two infinitesimally close states differ through the time dynamics of the system. Suppose there is a system with the state function $F_t(x)$, which gives the state after $t$ step starting from state $x$. Then Lyapunov exponent $\lambda$ is defined with an average over state space as:
\begin{equation}
    \left \langle F_t(x_0+\epsilon)-F_t(x_0) \right \rangle_{x_0} \approx \epsilon e^{\lambda t}
    \label{t_step_change}
\end{equation}
The exact value of the Lyapunov exponent can be found using the limit where the separation between states goes to 0.
\begin{align}
    \lambda t &\approx ln \left( \left \langle \frac{F_t(x_0+\epsilon)-F_t(x_0)}{\epsilon} \right \rangle_{x_0} \right )\\
    \lambda t &= \lim_{\epsilon \to 0}ln \left( \left \langle \frac{F_t(x_0+\epsilon)-F_t(x_0)}{\epsilon} \right \rangle_{x_0} \right )\\
    \lambda &=  \frac{1}{t} ln \left(  \left \langle \frac{dF_t(x)}{dx}  \right \rangle_{x_0}  \right )
    \label{lyapunov}
\end{align}

Analyzing Boolean networks with Lyapunov exponents is only possible by defining a Boolean derivative in place of the plain one in Eq.\ref{lyapunov}. In a Boolean network with $n$ nodes, the state is described with a $n$-dimensional Boolean vector. It is not viable to define infinitesimally close states in the Boolean setting, but by defining the distance between states as the number of nodes which is differed in states, the smallest distance between two states can be 1. The closest states only differ with only one entry. Suppose the state $x_0$ is defined as $x_0=(x_1,x_2,\dots,x_n)$ and then the average Boolean derivative of the one-step evolution function at this point is 
\begin{equation}
    \left \langle \frac{dF_1(x)}{dx}  \right \rangle_{x_0} =  \left \langle \sum\limits_{i=1}^n F(x_0^{i,1}) \oplus F(x_0^{i,0}) \right \rangle_{x_0} \ ,
    \label{network_sens}
\end{equation}
where $x_0^{i,1}$ is the state $x_0$ except $i^{th}$ entry is 1,  $x_0^{i,0}$ is the state $x_0$ except $i^{th}$ entry is 0, and $\oplus$ is the xor operation which gives the distance between two states. The expression in Eq.\ref{network_sens} is defined as the network sensitivity with the given Markovian time evolution function $F$.

One can also calculate the sensitivities of individual Boolean functions with the same expression. The sensitivity of a Boolean function $f$ is
\begin{eqnarray}
  \xi[f] &=& 2^{-n} \sum_{i=1}^n \sum_{\{x_0\}}f(x_0^{i,1}) \oplus
  f(x_0^{i,0})
  \label{sensitivity}
\end{eqnarray}
In this expression, the average over state space is given with another summation divided by the number of possible states. For a Boolean network, the average sensitivity of the Boolean functions of the nodes is equal to the network sensitivity defined in Eq.\ref{network_sens}.

The influence of a particular input can be also quantified with this formulation. The influence of the input $i$ on the Boolean function $f$ is
\begin{equation}
    \xi_i[f] =2^{-n} \sum_{\{x_0\}}f(x_0^{i,1}) \oplus f(x_0^{i,0}) 
    \label{influence}
\end{equation}
Quantitatively the influence is equal to the fraction of states in which flipping the $i^{th}$ input also flips the output of $f$. Subsequently, the sum of the influences of related inputs is equal to the sensitivity of the corresponding Boolean function.

\subsection{Criticality}

The chaos and stability can be distinguished with the value of the Lyapunov exponent $\lambda$. Considering Eq.\ref{t_step_change} positive values of $\lambda$ implies that a small deviation $\epsilon$ grows exponentially fast with time. Thus, a positive Lyapunov exponent occurs in a chaotic system, where negligible changes in the system constitute uncontrollable changes. On the contrary, negative values of the exponent result in an exponentially diminishing perturbation. Stable systems have a negative Lyapunov exponent and this leads to a critical point where the exponent is zero. When the exponent is zero, a system is neither chaotic nor stable. Instead, perturbations in the system flow with the same magnitude over time.

The stability condition of sensitivity values is hidden in the definition of sensitivity. Since sensitivity is actually the Boolean derivative in Eq.\ref{network_sens}. The expression of network sensitivity in terms of the Lyapunov exponent is,
\begin{equation}
    \xi = e^\lambda \ .
\end{equation}
Then the critical value of sensitivity is 1, and below it leads to stable time dynamics, above it is chaotic.

\section{Theoretical Discussion on Sensitivity and 2-Symmetry}

The sensitivity of $2-$symmetric functions can be written as a function of activity ratio of them. It is crucial to observe that the influence of symmetric inputs on Boolean functions is the same. 
\begin{eqnarray}
    \left \langle\frac{\partial f(\Vec{x})}{\partial x_i} \right \rangle_{\vec{x}} = \left \langle\frac{\partial f(S_{ij}\Vec{x})}{\partial x_j} \right \rangle_{\vec{x}} = \left \langle\frac{\partial f(\Vec{x})}{\partial x_j} \right \rangle_{\vec{x}} \ .
    \label{eq:sisj}
\end{eqnarray}

With this simplification, the influences of two inputs from different symmetry classes are sufficient to calculate sensitivity.

\begin{eqnarray}
    \xi_1 &=& 2^{-n} \sum_{\{x_0\}}f(x_0^{1,1}) \oplus f(x_0^{1,0}) \nonumber \\
     &=&  \frac{1}{2^n} \sum_{\Vec{x}} (g_2 \vee g_1(1,x_2, \dots, x_a) ) \oplus (g_2 \vee g_1(0,x_2, \dots, x_a) ) \nonumber\\
    &&\text{ By using the identity} (a\vee b) \oplus (a \vee c) = \bar{a} \wedge ( b \oplus c), \nonumber \\
    &=& \frac{1}{2^n} \sum_{\Vec{x}} \overline{g_2(x_{a+1}, \dots ,x_n)} \wedge  \frac{\partial g_1}{\partial x_1} \ \nonumber\\
\end{eqnarray}
If $\overline{g_2}=\overline{x_{a+1}} \wedge \dots \wedge \overline{x_n}$, then the summand is all connected with "and" operators. The summand is equal to 1 only when all terms are equal to 1 and there is only one configuration for the states $(x_{a+1},x_{a+2},\dots,x_n)=(0,0,\dots,0)$. For totally symmetric functions, the derivative with respect to any input is equal to 1 in only two states. A totally symmetric function is either connected with "or"s or "and"s; thus, there is only one state, all $1$s or $0$s, which gives the marginal output. For example, if the function is connected with "or"s then the marginal state is $(0,0,\dots,0)$. If an input is flipped then the output is also flipped. The other state, which equalizes the derivative with respect to the first input to 1, is the $(1,0,\dots,0)$. Thus, the summand is equal to 1 for only two states in the state space and the influence is equal to $\frac{2}{2^n}$.

In the other case, $\overline{g_2}=\overline{x_{a+1}} \vee \dots \vee \overline{x_n}$, there is only one state configuration $(x_{a+1},x_{a+2},\dots,x_n)$ which gives 0. Thus, there are $2^{n-a}-1$ states that set the first term to 1. Again, there are two states that set the derivative to 0. Then, the influence is $\frac{(2^{n-a}-1)2}{2^n}$
\begin{eqnarray}
     \xi_1&=&
    \begin{cases}
    2^{1-n} \textbf{ }\textbf{ }\textbf{ } \textbf{ }\textbf{ }\textbf{ } \textbf{ }\textbf{ }\textbf{ } \textbf{ }\textbf{ }\textbf{ },\text{ if } \overline{g_2}=\overline{x_{a+1}} \wedge \dots \wedge \overline{x_n} \\
    (2^{n-a}-1)2^{1-n}, \text{ if } \overline{g_2}=\overline{x_{a+1}} \vee \dots \vee \overline{x_n} \ .
    \end{cases}
\end{eqnarray}
Using a similar approach, one can also calculate $\xi_n$ as
\begin{eqnarray}
    \xi_n &=&
    \begin{cases}
    2^{1-n} \textbf{ }\textbf{ }\textbf{ } \textbf{ }\textbf{ }\textbf{ } \textbf{ }\textbf{ }\textbf{ }, \overline{g_1}=\overline{x_{1}} \wedge \dots \wedge \overline{x_a} \\
    (2^{a}-1)2^{1-n}, \overline{g_1}=\overline{x_{1}} \vee \dots \vee \overline{x_a}\ .
    \end{cases}
\end{eqnarray}
Finally, the sensitivity coefficients can be calculated exactly for all four choices of a 2-symmetric function $f$ by summing the influences of inputs.
\begin{eqnarray}
    \xi(n,a) = 
    \begin{cases}
    n 2^{1-n} &,\ \text{Type 0}\\
    a 2^{1-n}+(n-a) (2^{a}-1) 2^{1-n}&
    ,\ \text{Type 1} \\
    a (2^{n-a}-1) 2^{1-n}+(n-a) 2^{1-n}&
    ,\ \text{Type 2} \\
    a (2^{n-a}-1) 2^{1-n} +(n-a) (2^{a}-1) 2^{1-n}&
    ,\ \text{Type 3} \ .
    \end{cases}
    \label{eq:2-sensitivities}
\end{eqnarray}

By exploiting the logical structure one can also calculate the activity of a $2-$symmetric function. In the simplest case, the type 0 function has either $1$ or $2^n-1$ 1s in its truth table. Thus its activity ratio is either $2^{-n}$ or $1-2^{-n}$.

For type 1 functions $(f= (x_1 \wedge \dots \wedge x_a) \vee x_{a+1} \vee \dots \vee x_n)$ one can count the number of states which leads to 0 output. For 0 output from $x_{a+1}$ to $x_n$ and $(x_1 \wedge \dots \wedge x_a)$ must be 0. The number of states, which makes the term with parentheses 0, is equal to $2^a-1$. Thus the activity ratio of the function equals $1-(2^a-1)2^{-n}$ or $(2^a-1)2^{-n}$.

For type 2 functions it is enough to replace $a$ with $n-a$ in the activity ratio of type 1 functions.

For having a 0 output from type 3 functions $f= (x_1 \wedge \dots \wedge x_a) \vee (x_{a+1} \wedge \dots \wedge x_n)$, the terms with inside the parantheses must be 0. For the first one, there are $2^a-1$ states and for the second one, there are $2^{n-a}-1$ states. Together there are $(2^a-1)(2^{n-a}-1)$ many states which lead to 0 output. One can finally write the activity ratios of each type of function as:
\begin{align}
\alpha =
    \begin{cases}
    2^{-n} \text{ or } 1-2^{-n}&, \text{Type 0} \\
    (2^{a}-1)2^{-n} \text{ or } 1-(2^{a}-1)2^{-n}&, \text{Type 1}\\
    (2^{n-a}-1)2^{-n} \text{ or } 1-(2^{n-a}-1)2^{-n}&, \text{Type 2}\\
    (2^{n-a}-1)(2^{a}-1)2^{-n} \text{ or } 1-(2^{n-a}-1)(2^{a}-1)2^{-n}&,\text{Type 3} \\
    \end{cases}
\end{align}
Consequently, sensitivity $\xi$ can now be written as a function of activity ratio $\alpha$ as
\begin{eqnarray}
    \xi(\alpha^*)= 
    \begin{cases}
    -2\log_2(\alpha^*)\alpha^*&, \ \text{Type 0}\\
    2aK\alpha^*-2\alpha^* \log_2(\frac{K}{L}\alpha^*)&, \ \text{Type 1}\\
    2a\alpha^*-2 \log_2(1-2^a\alpha^*) (L-\alpha^*)&, \ \text{Type 2} \\
    2aK+2\log_2(-K+\frac{K}{L}\alpha^*)(L-\alpha^*)-2a\alpha^*K&,\ \text{Type 3}\ ,
    \end{cases}
\end{eqnarray}
where $\alpha^*=\min(\alpha,1-\alpha)$, $K=\frac{1}{2^a-1}$, and $L=\frac{1}{2^a}$. Note that all sensitivity coefficients are symmetric functions of the activity bias. From these expressions, the maximum possible sensitivity for each \textit{2-symmetric} function class can be obtained as
\begin{eqnarray}
    \xi^{max}=
    \begin{cases}
    1, \quad \text{for } (n,a)=(1,1) \text{ and } f=x_1 \vee \dots \vee x_n\\ 
    1.25, \quad \text{for } (n,a)=(3,1)
    \quad\text{and } f= (x_1 \wedge \dots \wedge x_a) \vee x_{a+1} \vee \dots \vee x_n \\
    1.25,\quad \text{for } (n,a)=(3,1)
    \quad\text{and } f= x_1 \vee \dots \vee x_a \vee (x_{a+1} \wedge \dots \wedge x_n) \\
    1.5, \quad \text{for } (n,a)=(4,2) 
    \quad\text{and } f= (x_1 \wedge \dots \wedge x_a) \vee (x_{a+1} \wedge \dots \wedge x_n)\ . &
    \end{cases}
    \label{eq:max_sensitivities}
\end{eqnarray}

In that sense, the symmetry condition has a limiting effect on the sensitivity and promotes stability while also favoring criticality since the maximum values are near the marginal sensitivity 1. 

\subsection{Limiting cases}

Observing the behavior of sensitivity when the input number increases, sensitivity increases linearly with input size for general Boolean functions $E[\xi]=n/2$ and for canalizing functions  $E[\xi]=(n+1)/4$. On the contrary, if we analyze the limits of equations (1) and (2), we encounter totally different characteristics.

For the case $f=x_1 \vee \dots \vee x_n$ (Type 0),
\begin{align*}
    \xi=\frac{a}{2^{n-1}}+\frac{(n-a)}{2^{n-1}} = \frac{n}{2^{n-1}} \\
    \alpha= 2^{-n}\\
    \text{Considering } 2^n >> n \\
    \xi\rightarrow 0, \alpha \rightarrow 0
\end{align*}\\

For the case $f= (x_1 \wedge \dots \wedge x_a) \vee x_{a+1} \vee \dots \vee x_n$ (Type 1),
\begin{align*}
    \xi=\frac{a}{2^{n-1}}+\frac{(n-a)(2^a-1}{2^{n-1}} = \frac{2a-n}{2^{n-1}}+\frac{2(n-a)}{2^{n-a}} \\
    \alpha= (2^a-1)2^{-n}\\
    \text{Considering } 2^n >> n \\
    \xi\rightarrow 0, \alpha \rightarrow 0
\end{align*}\\

For the case $f= x_1 \vee \dots \vee x_a \vee (x_{a+1} \wedge \dots \wedge x_n)$ (Type 2),
\begin{align*}
    \xi=\frac{a(2^{n-a}-1)}{2^{n-1}}+\frac{(n-a)}{2^{n-1}} = \frac{n-2a+a(2^{n-a})}{2^{n-1}}= \frac{2n-4a}{2^n}+\frac{2a}{2^a} \\
    \alpha= (2^{n-a}-1)2^{-n}=\frac{1}{2^a}-\frac{1}{2^n}\\
    \text{Considering } 2^n >> n \\
    \xi\rightarrow \frac{2a}{2^a}, \alpha \rightarrow \frac{1}{2^a}
\end{align*}\\

For the case $ f= (x_1 \wedge \dots \wedge x_a) \vee (x_{a+1} \wedge \dots \wedge x_n)$ (Type 3),
\begin{align*}
    \xi=\frac{a(2^{n-a}-1)}{2^{n-1}}+\frac{(n-a)(2^a-1)}{2^{n-1}} =  \frac{-2n}{2^n}+\frac{2a}{2^a}+\frac{2(n-a)}{2^{n-a}} \\
    \alpha= (2^{n-a}-1)(2^a-1)2^{-n}=1-\frac{1}{2^a}-\frac{2^a-1}{2^n}\\
    \text{Considering } 2^n >> n, 2^{(n-a)}>> n \\
    \xi \rightarrow \frac{2a}{2^a}, \alpha \rightarrow \frac{1}{2^a}
\end{align*}
\begin{figure}[!h]
  \centering
  \begin{minipage}[b]{0.47\textwidth}
    \includegraphics[width=\textwidth]{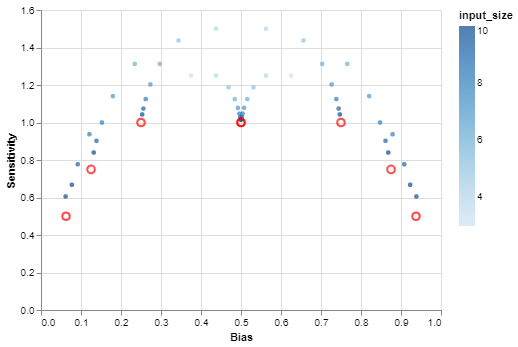}
    \caption{Sensitivity vs. Activity bias of Type-3 functions}
  \end{minipage}
  \hfill
  \begin{minipage}[b]{0.47\textwidth}
    \includegraphics[width=\textwidth]{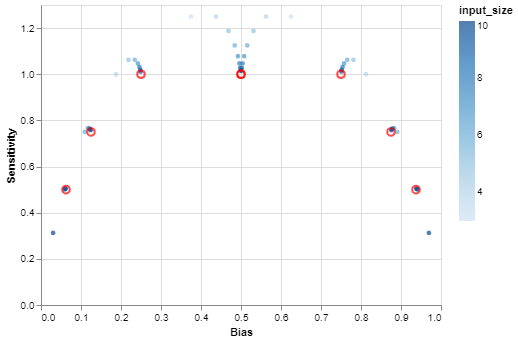}
    \caption{Sensitivity vs. Activity bias of Type-2 functions}
  \end{minipage}
\end{figure}

\begin{figure}[!h]
  \centering
    \includegraphics[width=0.6\textwidth]{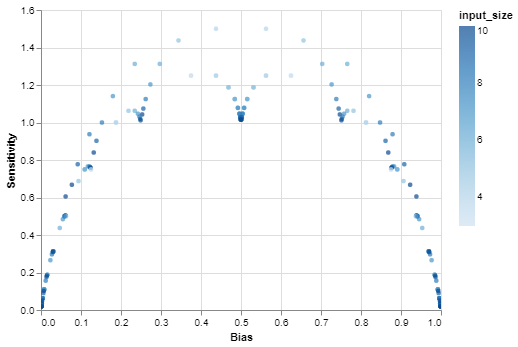}
    \caption{Sensitivity vs. Activity bias of all functions with at most 2 symmetry classes}
\end{figure}

\section{Theoretical Discussion on Sensitivity of Canalizing Functions}

\subsection{The relation between canalization and expected sensitivity}
Few of the studies on canalization and sensitivity in the Boolean context illustrate a quantitative effect of canalization on sensitivity. It is known that for a Boolean function with an activity ratio $p$ and the number of inputs $n$, the expected sensitivity is $2np(1-p)$. It is simply followed from Eq.\ref{sensitivity}. Given an activity ratio, the terms in the equation differ from each other with a probability $2p(1-p)$. Then the summation over inputs gives the expected sensitivity value of $E[\xi]=2np(1-p)$. The expected sensitivity of a canalizing function is given as $(n+1)/4$ assuming the expected value of activity ratio is $1/2$ \cite{shmulevich2004activities}. Knowing that there are several canalizing inputs, the expectation of sensitivity for a partially canalizing function with a canalizing depth $k$ is $1+(n-k-2)/(2^{k+1})$ \cite{layne2012nested}. This expression includes the first two cases: random and random canalizing ones, where $k=0$ and $k=1$, respectively.

However, knowing the activity ratio of a Boolean function should also affect the expectations. Thus, in this section, we will discuss the expected sensitivities of canalizing functions in relation to the mean activity ratio. We derive a direct function of activity ratio for the exact sensitivity of a NCF and this function forms the minimum-sensitivity curve for Boolean functions. The function also leads to an expression of the expected sensitivities of partially canalizing functions.

\subsection{The fractal relation between activity ratio and sensitivity of NCFs}

Sensitivity can also be inspected using the hypercube representation of Boolean functions. A Boolean function with $n$ inputs can be described as $n$-dimension hypercube since this cube has $2^n$-many vertices. Each vertex corresponds to an input state and the color of the vertex (filled for 1, empty for 0) gives the output of the function given an input state.  In this representation, there is always an edge between states which differs with only one entry like in the Eq.\ref{sensitivity}. Thus, sensitivity can be calculated by counting the boundary edges which are the edges connecting differently-colored vertices. The expression for the sensitivity is 
\begin{equation}
    \xi[f]=\frac{2b[f]}{2^{n}} \ ,
\end{equation}
where $b[f]$ is the number of boundary edges in the hypercube representation of $f$.

A subsequent observation is that sensitivity is invariant under complement operation since boundary edges do not change. Assuming the activity ratio $p$ is then $1/2$ does not impair the generality. Suppose there is a nested canalization with an activity ratio $p$ less than half. Thus, immediately this means that $r_1=0$ by the Eq.\ref{ncf_p}. Also, the activity is then given with the expression, 
\begin{eqnarray}
    \alpha = \sum\limits_{i=2}^n r_i\times 2^{i-1}
\end{eqnarray}

Then, this activity is equal to the NCF which is obtained by setting the first canalizing variable $x_1$ to the inverse of its canalizing value $\sigma_1$. Visually the activity of a function is the number of filled states in its hypercube representation. The reduction process of an NCF corresponds to the removal of a hyper-face, which consist of only empty vertices since the activity of the NCF and reduced version is the same. Comparing the boundary edges of these functions, the only difference originates from the empty (or uniform) hyper-face and its counterpart, which has the $\alpha$ filled  vertices. Thus, in between them, there are $\alpha$ boundary edges. Then, in an expression, this relation can be given as:
\begin{eqnarray}
    \beta(n,\alpha)=\beta(n-1,\alpha)+\alpha \ , \ \alpha < 1/2
    \label{beta_reduced}
\end{eqnarray}

\begin{figure}[h!]
    \centering
    \includegraphics[width=0.8\textwidth]{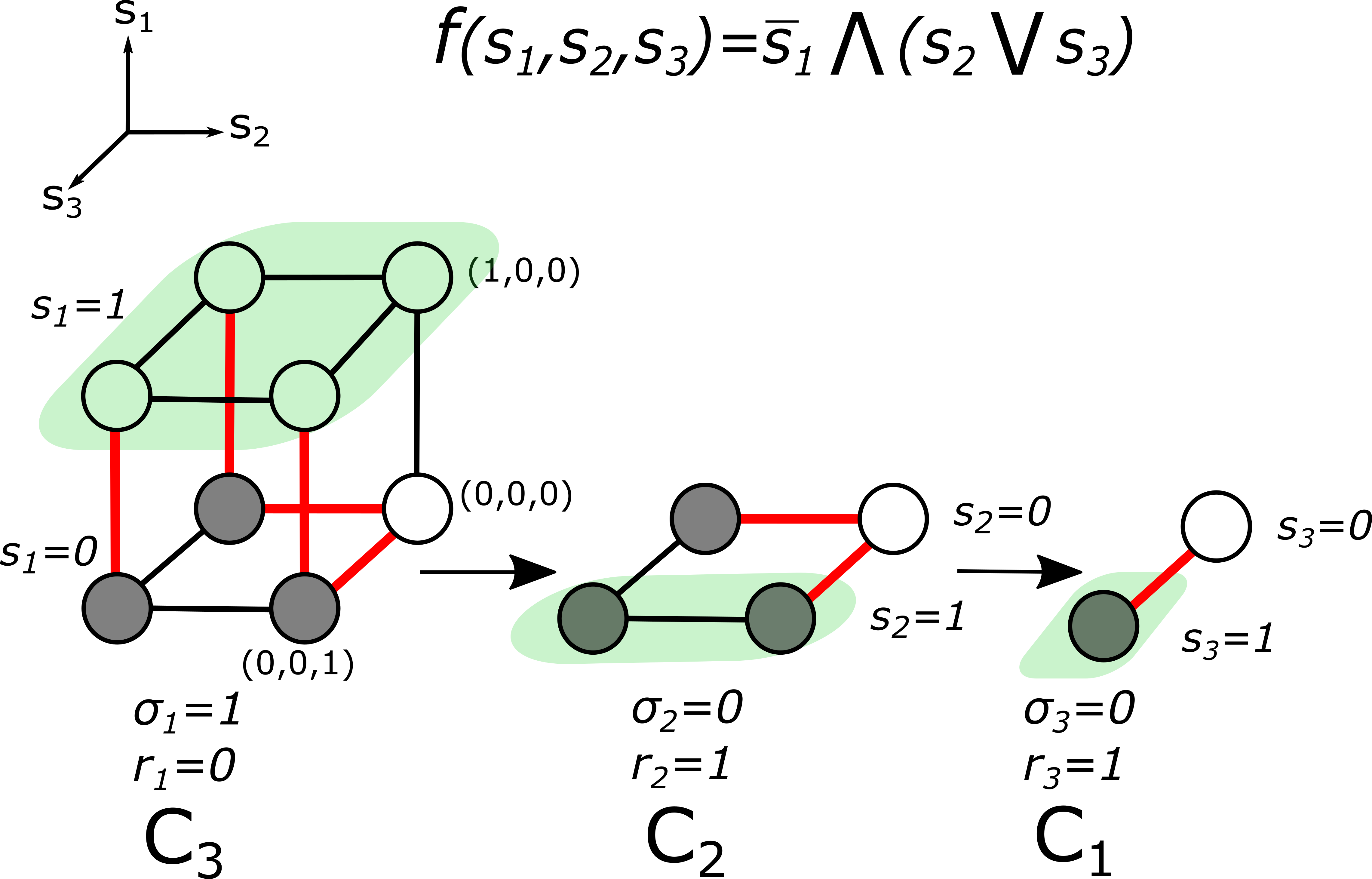}
    \caption{The hypercube representation of a simple nested canalizing function. Uniform hyper-faces are shown in the green shade. In each step, the NCF is reduced by removing the uniform hyper-face.}
    \label{fig:hypercube}
\end{figure}

where $\beta(n,\alpha)$ gives the number of boundary edges of an NCF with $n$ inputs and an activity $\alpha$. The invariance of sensitivity under complement operation follows the equation:
\begin{eqnarray}
    \beta(n,\alpha)=\beta(n,2^n-\alpha)
    \label{beta_negation}
\end{eqnarray}

An NCF may include redundant inputs which constitute redundant state separations. If redundantly separated state pairs are merged together, the number of inputs can be reduced by
\begin{eqnarray}
    \beta(n,\alpha)=2\beta(n-1,\alpha/2)  \ .
    \label{ncf_even}
\end{eqnarray}

\subsection{The sensitivity as a function activity ratio}

By using Eq.\ref{beta_negation} and \ref{beta_reduced} it is possible to relate the number of boundary edges of any NCF with the simplest Boolean functions with one input. The algorithm for finding the boundary edges of an NCF consists of two cases. 

\begin{enumerate}
    \item If $\alpha < 1/2$, reduce the number of inputs with Eq.\ref{beta_reduced}. $\beta \xrightarrow{} \beta+\alpha, \ n\xrightarrow{}n-1$\\
    \item Else, use the Eq.\ref{beta_negation} to make the activity less than $1/2$. Then, apply the first step.
    \item Stop if $n=0$.
\end{enumerate}

The algorithm finds the boundary edges of an NCF with a given activity. Since the activity ratio uniquely defines the pair $(n,p)$, this leads to a functional relation between sensitivity and the activity ratio of a Boolean function. The sensitivity of an NCF as a function of its activity ratio is given with
\begin{eqnarray}
    \xi_{NCF}(p)=\frac{\beta(n,\alpha)}{2^{n-1}} \ .
     \label{xi_def}
\end{eqnarray}
By dividing both sides to $2^{n-1}$ and setting $p=\alpha/2^{n}$, Eq.\ref{beta_reduced} and Eq.\ref{beta_negation}  become
\begin{align}
    \xi_{NCF}(p)&=\frac{\xi_{NCF}(2p)}{2}+2p \ , \ p\leq 1/2 \label{xi_recursive}\\
    \xi_{NCF}(p)&=\xi_{NCF}(1-p) \label{xi_negation}\\
\end{align}

\begin{figure}[h!]
    \centering
    \includegraphics[width=0.6\textwidth]{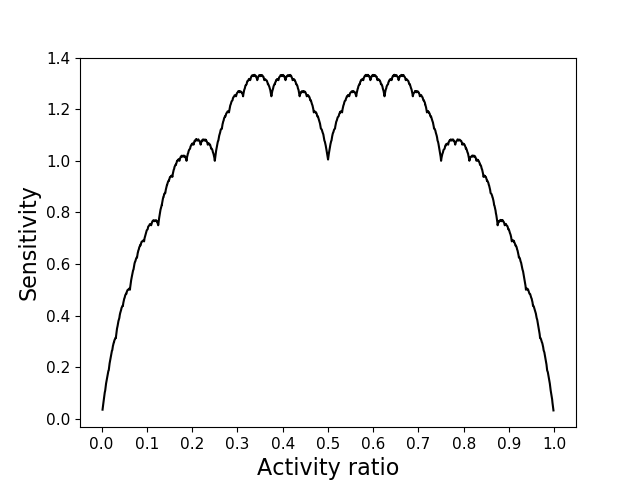}
    \caption{The complete curve of the function $\xi_{NCF}(p)$. (Known as the Blancmange curve or Takagi curve)}
    \label{fig:blancmange}
\end{figure}

Assuming a uniform activity ratio distribution the expected sensitivity of an NCF is exactly 1. In this regard, NCFs tend to organize a critical behavior if there is no pressure on the choice of activity ratio. 

The equations \ref{xi_recursive} and \ref{xi_negation} correspond to the definition of the Takagi curve, which is also known as the Blancmange curve in Fig.\ref{fig:blancmange} This curve is one of the most studied fractal functions, and it is known as continuous everywhere, but nowhere differentiable. This curve is also encountered in other graph theory-related problems \cite{GUU2000163,FRANKL1995125}. 

\subsection{The minimum curve for Boolean functions}

Since in numerical analysis, it is impossible to create less sensitive functions than NCFs for a given activity ratio, the Blancmange curve is suspected to be the minimal sensitivity boundary for all Boolean functions. The objective is to show that the number of boundary edges of NCFs, which is given with the function $\beta$, is the minimum among the functions with the same number of inputs and activity. 

\begin{eqnarray}
  \beta(n,\alpha) &=& \min_{f\in {\cal B}_{n,\alpha}} b[f],\ \forall  n,\alpha \ ,
                 \label{objective}
\end{eqnarray}
where ${\cal B}_{n,\alpha}$ is the set of Boolean functions with $n$ inputs and activity $\alpha$. It is convenient to restate the objective such that it includes an inequality. 

\begin{eqnarray}
  \beta(n,\alpha) &\leq&  b[f],\ \forall  n,\alpha \ , \text{ and} \ \forall f \in {\cal B}_{n,\alpha} 
                 \label{objective}
\end{eqnarray}
The minimality property can be rephrased by using the hypercube graphs as follows. The objective is to show that there is no coloring on the hypercube using $\alpha$-many fillings such that the coloring has fewer boundary edges than the NCF counterpart. The coloring scheme can be divided into two hyperfaces of the graph, which both correspond to Boolean functions with one less input. Then, the activity $\alpha$ of a Boolean $f$ is shared between Boolean functions (or hyper-faces) $f_1$ and $f_2$. If the activity of $f_1$ is given with $\alpha_1$, $f_2$ has the activity of $\alpha-\alpha_1$. Since the number of boundary edges between these hyper-faces is at least the difference in activities between them the objective is to prove  
\begin{equation}
 \beta(n,\alpha) \leq b[f_1]+b[f_2] + |\alpha-2\alpha_1| \ ,\ \forall f_1
 \in {\cal B}_{n-1,\alpha_1}\ \mbox{and}\ \ \forall f_2 \in {\cal
   B}_{n-1,\alpha-\alpha_1} \ .
\label{general_ineq}
\end{equation}
Since the set of all Boolean functions with two inputs is small it can be shown that Ineq. \ref{general_ineq} holds for $n=2$. The rest of the proof is based on induction and by the inductive hypothesis 
\begin{eqnarray}
     \beta(n-1,\alpha_1)+\beta(n-1,\alpha-\alpha_1)+ |\alpha-2\alpha_1|\leq b[f_1]+b[f_2] + |\alpha-2\alpha_1| \  
\end{eqnarray}
follows. It is sufficient to show that the inequality
\begin{eqnarray}
     \beta(n,\alpha)\leq \beta(n-1,\alpha_1)+\beta(n-1,\alpha-\alpha_1)+  |\alpha-2\alpha_1|, \ \forall  n,\alpha
\label{main}
\end{eqnarray}
holds.
By the induction hypothesis
\begin{equation}
    \begin{aligned}
         \beta(n-1,\alpha) \leq \beta(n-2,\alpha_1)+\beta(n-2,\alpha-\alpha_1)+| \alpha-2\alpha_1 |
    \end{aligned}
    \label{induction_hyp}
\end{equation}
is true for all $\alpha$ and $\max(0,\alpha-2^{n-2}) \leq \alpha_1\leq
\min(\alpha,2^{n-2})$. Note that, it is sufficient to consider $\alpha<2^{n-1}$
by virtue of Eq.(\ref{beta_negation}).

For rigorous proof, we should carefully utilize the properties of the
function $\beta()$ by separating the problem into four regions
corresponding to different values of $\alpha$ and $\alpha_1$.

\begin{figure}[h!]
    \centering
    \includegraphics[width=0.5\linewidth]{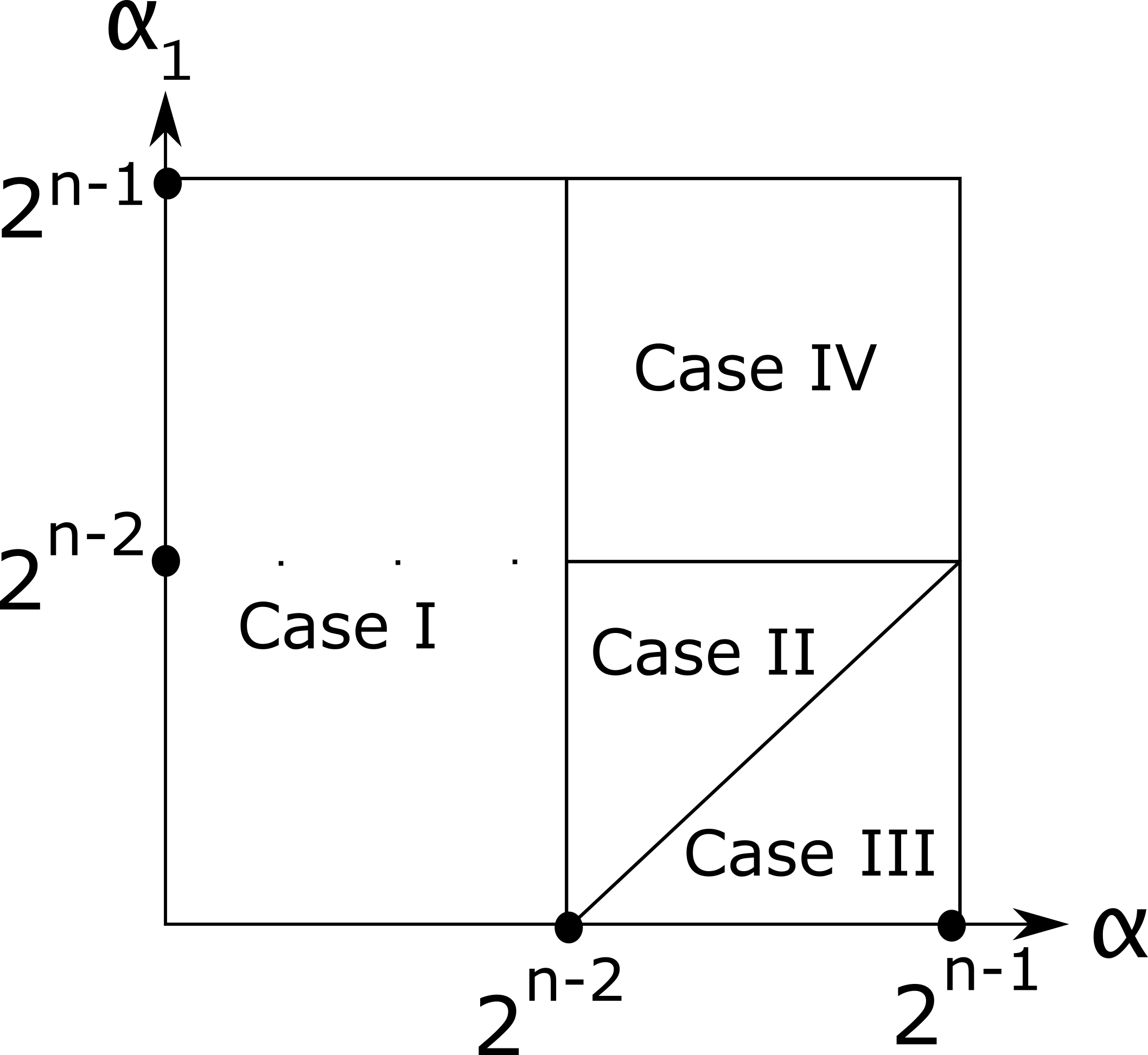}
    \caption{Proof by induction is done separately for the shown four
      regions in $(\alpha,\alpha_1)$-plane.}
    \label{fig:proof_cases}
\end{figure}

\underline{Case I.} When $\alpha<2^{n-2}$, substituting Eq.(\ref{beta_reduced}) in
Ineq.(\ref{induction_hyp}) immediately yields the desired expression:
\begin{align*}
 \beta(n,\alpha) \leq \beta(n-1,\alpha_1)+\beta(n-1,\alpha-\alpha_1)+| \alpha-2\alpha_1 |
\end{align*}
for all $\alpha$ and $0 \leq \alpha_1\leq \alpha$.

\underline{Case II.} When $\alpha> 2^{n-2}$, the substitution above
still works if $\textbf{ } \alpha-2^{n-2}\leq \alpha_1\leq 2^{n-2}$,
however, more work is needed for values $\alpha_1$ lying above and
below this region in Fig.(\ref{fig:proof_cases}).

\underline{Case III.} We first consider the region where $0\leq
\alpha_1\leq \alpha-2^{n-2}$. Now write the inductive hypothesis in a
slightly different form as
\begin{eqnarray}
     \beta(n-1,\alpha-2^{n-2}) &\leq& \beta(n-2,\alpha_1) +\beta(n-2,\alpha-2^{n-2}-\alpha_1) \nonumber \\
     && +| \alpha-2^{n-2}-2\alpha_1 |
\label{extended_region}
\end{eqnarray}
for all $h$ and $0\leq \alpha_1\leq \alpha-2^{n-2}$, and manipulate it
using Eqs.(\ref{beta_negation},\ref{beta_reduced}):
 \begin{eqnarray}
     \beta(n-1,\alpha-2^{n-2})&=&\beta(n-2,\alpha-2^{n-2})+\alpha-2^{n-2} \nonumber\\
     &=& \beta(n-2,2^{n-1}-\alpha)+\alpha-2^{n-2}\nonumber \\
     &=& \beta(n-1, 2^{n-1}-\alpha)+\alpha-2^{n-2}+\alpha-2^{n-1}\nonumber \\
     &=& \beta(n-1,\alpha) +2\alpha -3 \times2^{n-2} \nonumber \\
     &=& \beta(n,\alpha)+\alpha - 3 \times 2^{n-2}\ .
     \label{connection}
 \end{eqnarray}
Substituting in Ineq.(\ref{extended_region}), we obtain
 \begin{flalign}
    &\beta(n,\alpha)+\alpha - 3 \times 2^{n-2}
   \leq \beta(n-2,\alpha_1) +\beta(n-2,\alpha-2^{n-2}-\alpha_1) +| \alpha-2^{n-2}-2\alpha_1 | \nonumber \\
    &\beta(n,\alpha)+\alpha - 3 \times 2^{n-2} \leq \beta(n-1,\alpha_1)-\alpha_1+\beta(n-2,2^{n-1}-\alpha+\alpha_1) +| \alpha-2^{n-2}-2\alpha_1 |\nonumber \\
   &\leq  \beta(n-1,\alpha_1)-\alpha_1+\beta(n-1,2^{n-1}-\alpha+\alpha_1)-2^{n-1}+\alpha-\alpha_1+| \alpha-2^{n-2}-2\alpha_1 |\nonumber \\
   &\mbox{yielding,} \nonumber \\
   &\beta(n,\alpha) \leq\beta(n-1,\alpha_1) + \beta(n-1,\alpha-\alpha_1)+2^{n-2}-2\alpha_1+| \alpha-2^{n-2}-2\alpha_1 |.
     \label{case1}
 \end{flalign}
In order to reach Ineq.(\ref{main}) we need to further show
that
\begin{equation}
  2^{n-2}-2\alpha_1+| \alpha-2^{n-2}-2\alpha_1| \leq |\alpha -2\alpha_1|.
  \label{intermediate}
\end{equation}\\
First observe that, since $2^{n-2} < \alpha < 2^{n-1}$ and $0\leq
\alpha_1 \leq \alpha-2^{n-2}$, we have $\alpha \geq
2\alpha_1$.\\ \\ There are two possibilities. If $\alpha-2^{n-2} \geq
2\alpha_1$,
\begin{align*}
  2^{n-2}-2\alpha_1+| \alpha-2^{n-2}-2\alpha_1 |= \alpha- 4\alpha_1 \\
  \leq | \alpha -2\alpha_1| = \alpha-2\alpha_1.
\end{align*}
Otherwise $\alpha-2^{n-2} \leq 2\alpha_1$ and again,
\begin{align*}
    2^{n-2}-2\alpha_1+| \alpha-2^{n-2}-2\alpha_1 |= 2^{n-1}-\alpha\\
    \leq |\alpha-2\alpha_1|,\ \ \text{ since }\ \alpha_1\leq \alpha-2^{n-2}.
\end{align*}
Having shown that Ineq.(\ref{intermediate}) holds, we substitute it in
Ineq.(\ref{case1}) to reach the desired results for $0\leq \alpha_1\leq
\alpha-2^{n-2}$:\\
 \begin{eqnarray}
   \beta(n,\alpha) &\leq& \beta(n-1,\alpha_1) + \beta(n-1,\alpha-\alpha_1) +2^{n-2}-2\alpha_1+| \alpha-2^{n-2}-2\alpha_1 |\nonumber \\
   &\leq&  \beta(n-1,\alpha_1) + \beta(n-1,\alpha-\alpha_1)+| \alpha-2\alpha_1 |,\ \textbf{ }\forall \alpha,\alpha_1 \textbf{ }  0\leq \alpha_1\leq \alpha-2^{n-2}. \nonumber
 \end{eqnarray}
\underline{Case IV.} Finally consider the region,  where
$2^{n-2}\le \alpha_1 \le \alpha$ and define $\alpha_1'=\alpha_1+2^{n-2}$. We express the
induction hypothesis as
\begin{eqnarray}
  \beta(n-1,\alpha-2^{n-2})
  &\leq& \beta(n-2,\alpha_1'-2^{n-2})+\beta(n-2,\alpha-\alpha_1') +| \alpha-2\alpha_1'+2^{n-2} | \nonumber
\end{eqnarray}
for all $\alpha$ and $\textbf{ } 2^{n-2}\leq \alpha_1'\leq h$. Substituting
Eq.(\ref{connection}) above, we obtain
\begin{eqnarray}
&&\beta(n,\alpha)+\alpha -3\times 2^{n-2} \leq \beta(n-2,2^{n-1}-\alpha_1')+\beta(n-2,\alpha-\alpha_1')+| \alpha-2\alpha_1'+2^{n-2} | \nonumber \\
  &\leq& \beta(n-1,2^{n-1}-\alpha_1')+\alpha_1'-2^{n-1}+\beta(n-1,\alpha-\alpha_1')+\alpha_1'-\alpha+| \alpha-2\alpha_1'+2^{n-2} | \nonumber
\end{eqnarray}
yielding,
\begin{eqnarray}
  &&\beta(n,\alpha) \leq \beta(n-1,\alpha_1')+\beta(n-1,\alpha-\alpha_1') +2\alpha_1'-2\alpha+2^{n-2}+| \alpha-2\alpha_1'+2^{n-2} |. \nonumber
 \end{eqnarray}
The final step is to show that
\begin{equation}
  2\alpha_1'-2\alpha+2^{n-2}+| \alpha-2\alpha_1'+2^{n-2} | \leq | \alpha-2\alpha_1'|
  \label{final_step}
\end{equation}
for $2^{n-2}<\alpha<2^{n-1}$ and $2^{n-2}<\alpha_1'<\alpha$. Since $|
\alpha-2\alpha_1'|=2\alpha_1'-\alpha$, Ineq.(\ref{final_step}) reduces to
\begin{equation}
  2^{n-2}+| \alpha-2\alpha_1'+2^{n-2} |\leq \alpha.
  \label{final_step2}
\end{equation}
When $\alpha-2\alpha_1'+2^{n-2} \geq 0$, Ineq.(\ref{final_step2}) follows from
$\alpha_1'\geq 2^{n-2}$ and otherwise from $\alpha_1'\leq \alpha$. Therefore, we find
that Ineq.(\ref{main}) is satisfied for $2^{n-1}\ge \alpha \ge 2^{n-2}$ as
well, which completes the proof.

On a peripheral note, there is a
correspondence between Boolean functions and the microstates of the
Ising model on a hypercube, through the coloring scheme described in
the text. By this analogy, the boundary edges are the excitations of
the corresponding Ising model. Therefore, the minimum sensitivity of
NCFs established here can also be interpreted as NCFs being the
ground-states of the Ising model on a hypercube at any given
magnetization. NCFs are also known in the realm of computer science -
under the alternate identity of ``unate cascade functions'' - due to
their optimal properties in the context of binary decision
processes~\cite{jarrah2007nested}.
\section{Theoretical Discussion of Sensitivity and Partially Nested Canalization}

The relation between the hypercube graph and sensitivity is already discussed in previous sections. In the case of partially nested canalizing functions (pNCF), the recursive relation in Eq. \ref{beta_reduced} ends prematurely and distorts the fractal relation between activity ratio and sensitivity. The distortion is proportional to the deviation from an NCF, i.e. if the canalization depth $k$ is closer to input number $n$, the distortion would be minimal. Additionally, the exact relation between sensitivity and activity ratio does not exist for pNCFs and the expectation of sensitivity as a function of activity ratio is plausible with a looser nature of pNCFs than NCFs.

Suppose $\langle \beta(n,k,\alpha) \rangle$ is the expected sensitivity for a pNCF with $n$ inputs, a canalization depth $k$, and an activity $\alpha$. Since the nested structure is present we can construct a relation similar to Eq. \ref{beta_reduced}.
\begin{equation}
    \langle\beta(n,k,\alpha)\rangle=\langle\beta(n-1,k-1,\alpha)\rangle+ \alpha \ ,\text{if} \ \alpha<2^{n-1} , \ k>1 \ .
    \label{pncf_reduced}
\end{equation}
The negation principle still works as
\begin{equation}
    \langle\beta(n,k,\alpha)\rangle=\langle\beta(n,k,2^n-\alpha)\rangle \ .
\end{equation}
However, the Eq.\ref{pncf_reduced} ends when $k=1$ with
\begin{equation}
    \langle\beta(n,1,\alpha[\hat{f}])\rangle=\langle\beta[\hat{f}]\rangle+ \langle\alpha[\hat{f}]\rangle \ ,
\end{equation}
If $\hat{f}$ would be an NCF, then the expectation exactly gives fractal relation for NCFs. Consequently, the only difference between pNCFs and NCFs comes from the expected structure of non-canalizing function $\hat{f}$. Then, it is valid to define the expected sensitivity of pNCFs as
\begin{equation}
     \langle\beta(n,k,\alpha)\rangle= \beta(n,\alpha)+\langle\beta[\hat{f}]\rangle-\beta(n-k,\alpha[\hat{f}])
\end{equation}
If we divide both sides by $2^{n-1}$, the expected sensitivity of pNCFs is revealed as,
\begin{equation}
    \langle\xi(n,k,p)\rangle=\xi_{NCF}(n,p)+\frac{\langle\xi[\hat{f}]\rangle-\xi_{NCF}(n-k,p[\hat{f}])}{2^k}
\end{equation}

Noticing the activity of $\hat{f}$ is determined by the activity of $f$ with Eq.\ref{pncf_activity}, the activity ratio of $\hat{f}$ is 
\begin{equation}
    p[\hat{f}]= 2^k p - \left \lfloor 2^k p \right \rfloor
\end{equation}
Since $\hat{f}$ is a random Boolean function with $n-k$ inputs and an activity ratio $p[\hat{f}]$, its activity ratio is expected to be $2(n-k)p[\hat{f}](1-p[\hat{f}])$. Utilizing this expectation and remembering the independence of an NCF's sensitivity from its input number, the expected sensitivity of a pNCF is exactly defined with three parameters 

\begin{equation}
    \langle\xi(n,k,p)\rangle=\xi_{NCF}(p)+\frac{2(n-k)p[\hat{f}](1-p[\hat{f}])-\xi_{NCF}(p[\hat{f}])}{2^k} \ .
    \label{pncf_sensitivity}
\end{equation}

The deviation from the perfect fractal curve is quantified in this expression and it is inversely proportional to $2^k$. The deviation decays rapidly, i.e. it is negligible for canalization depths $k$ sufficiently close to $n$. If $n-k$ and $2^k$ are comparable, deviations would be relevant. 

\begin{figure}[h!]
    \centering
    \includegraphics[width=0.8\textwidth]{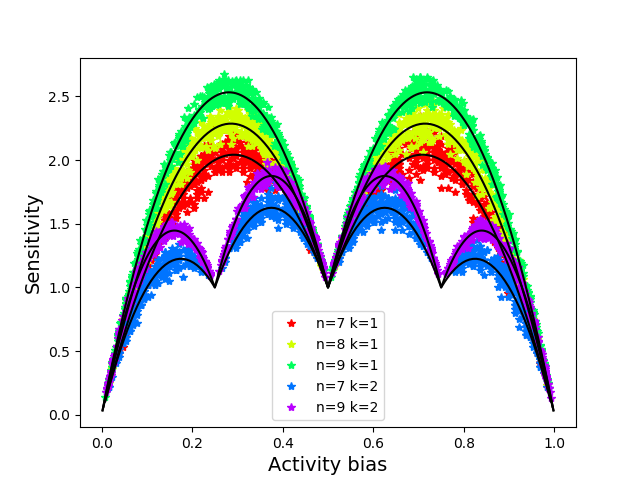}
    \caption{The sensitivity vs. activity bias of randomly generated pNCFs with different input numbers and canalization depths. The black lines show the expectation curve in Eq.\ref{pncf_sensitivity}}
    \label{fig:curve_pncf}
\end{figure}

Though the Eq.\ref{pncf_sensitivity} includes fractal terms, it does not constitute fractal behavior as depicted in Fig.\ref{fig:curve_pncf}. The combination of the fractal terms is guaranteed to have a non-fractal contribution as a consequence of Eqs. \ref{xi_recursive} and \ref{xi_negation}. The fractal terms combined is $\xi_{NCF}(p)-\xi_{NCF}(2^k p -\left \lfloor 2^k p \right \rfloor )/2^k$ and $\left \lfloor 2^k p \right \rfloor = \alpha \in \mathbb{N}$. If $\alpha$ is even, then the extension below is valid.
\begin{equation}
    \xi_{NCF}(2^k p - \alpha)=2\xi_{NCF}(2^{k-1} p - \alpha/2)-2^{k+1} p + 2\alpha \ .
    \label{pncf_extension}
\end{equation}
If $\alpha$ is odd, by the negation, the term can be converted to an even number, and the extending procedure is applicable until this mechanism reaches activity bias $p$. The activity bias is halved, and the sensitivity is doubled at each extending step. This mechanism should end after $k$ extending steps since the target activity bias is $p$. Consequently, after extensions, the fractal terms cancel each other the accumulation of non-fractal terms in Eq.\ref{pncf_extension} contributes to the average expected sensitivity of pNCFs.

For a nontrivial example $\alpha=1$, it implies $2^{-k}<p< 2^{-k+1}$ 
\begin{align*}
    \xi_{NCF}(2^k p - 1)&=\xi_{NCF}(2-2^{k} p)\\
    &=2\xi_{NCF}(1-2^{k-1}p)-4+2^{k+1} p\\
    &= 2\xi_{NCF}(2^{k-1}p)- 4+2^{k+1} p\\
    &= 2(2\xi_{NCF}(2^{k-2}p)-2^{k}p)-4+2^{k+1} p\\
    &= 4\xi_{NCF}(2^{k-2}p)-2^{k+1}p-4+2^{k+1} p\\
    &= 2^k\xi_{NCF}(p)-(k-1)2^{k+1}+p-4+2^{k+1} p\\
    &= 2^k\xi_{NCF}(p)-(k-2)2^{k+1}p-4
\end{align*}

So the expected sensitivity is equal to,
\begin{align*}
    \langle\xi(n,k,p)\rangle&=\xi_{NCF}(p)+\frac{2(n-k)p[\hat{f}](1-p[\hat{f}])-\xi_{NCF}(p[\hat{f}])}{2^k}\\
    &=\xi_{NCF}(p)+\frac{2(n-k)p[\hat{f}](1-p[\hat{f}])}{2^k}-\frac{ 2^k\xi_{NCF}(p)-(k-2)2^{k+1}p-4}{2^k}\\
    &=\frac{2(n-k)p[\hat{f}](1-p[\hat{f}])}{2^k}+\xi_{NCF}(p)-\xi_{NCF}(p)+\frac{4}{2^k}+2(k-2)p\\
    &=\frac{2(n-k)p[\hat{f}](1-p[\hat{f}])}{2^k}+\frac{4}{2^k}+2(k-2)p\\
    &=\frac{2(n-k)(2^kp-1)(2-2^kp)}{2^k}+\frac{4}{2^k}+2(k-2)p\\
    &=2(n-k)\left (-2^kp^2+3p-\frac{2}{2^k} \right) +\frac{4}{2^k}+2(k-2)p
\end{align*}    

The nested structure always ensures that for any given $n$ and $k$, the expected sensitivity of pNCFs consists of $2^k-$many piecewise explicit functions of $n,k$, and $p$. Unlike NCFs, the sensitivity curve of pNCFs does not possess fractal behavior. 

Assuming a uniform activity ratio distribution, the expected sensitivity of NCFs is exactly one. Comparably for pNCFs, it becomes 
\begin{equation}
    \langle\xi(n,k)\rangle=1+\frac{(n-k-3)}{2^k3}
\end{equation}
since the average of $p[\hat{f}](1-p[\hat{f}])$ is $1/6$. This expectation is more accurate for higher values of $n-k$. As well as $2-$symmetric and nested canalizing functions, pNCFs may promote criticality for sufficient canalization depths.

\section{Numerical Discussion}

Cell Collective database includes Boolean network models of real gene networks constructed by experiments. There are 76 networks, which consist of 3460 genes (Boolean functions) in total. Together with the theoretical study using the experimental data, further inspection of the criticality phenomenon and minimum sensitivity realization by NCFs is provided.

\subsection{General Characteristics of Experimental Data}
\begin{figure}[h!]
    \centering
    \includegraphics[width=0.6\textwidth]{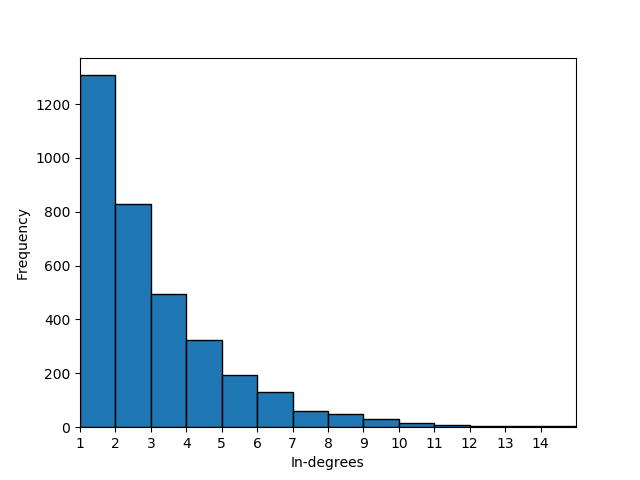}
    \caption{The in-degree distribution of genes in the Boolean network models of GRNs acquired from CellCollective.}
    \label{fig:in_degree}
\end{figure}

The input number of the Boolean functions, i.e. the connectivity of a gene, is a factor on its sensitivity. For a random Boolean function the expected sensitivity scales linearly with the number of inputs. More than one-third of the Boolean functions in the database have only single input. The sensitivity of such functions is always 1; thus, they are not interesting targets for sensitivity analysis. The whole input (in-degree) distribution can be observed in Fig. \ref{fig:in_degree}.

\begin{figure}[h!]
    \centering
    \includegraphics[width=0.8\textwidth]{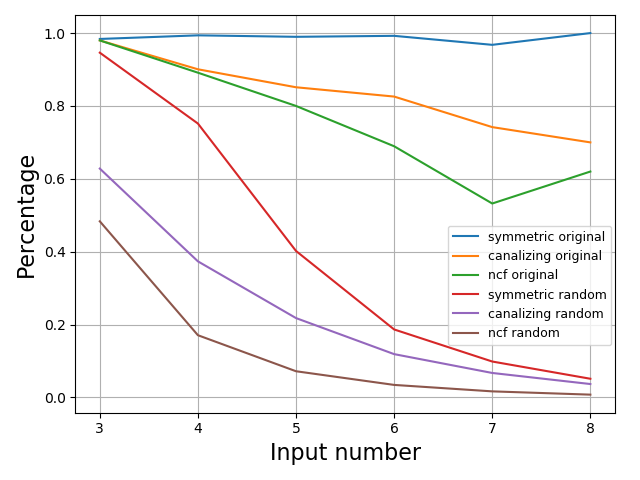}
    \caption{The fraction of canalizing, symmetric, and nested canalizing functions with different input numbers observed in biological networks and randomly generated ones. Original refers to biological.}
    \label{fig:freq_feat}
\end{figure}

The previous chapter discusses the effect of canalization and symmetry on sensitivity due to the abundance of these features in biological networks. These networks are inclined to acquire these features radically more than their random counterparts, as seen in Fig. \ref{fig:freq_feat}. The probability of possessing symmetry or canalization drops significantly with increasing input numbers. Random Boolean functions are generated by choosing an activity ratio from a uniform distribution, and the states which give the output, 1, are chosen randomly. 

The overall fractions of special functions in these networks are given in the table \ref{tab:features} by ignoring functions with less than $3$ inputs since they are always symmetric and nested canalizing. Though symmetry is a dominant feature, its effect on sensitivity is unclear except for $2-$ symmetric functions. The relation between nested canalization and sensitivity is fully discovered, and the average sensitivity of an NCF network (or its criticality) only depends on the activity ratio distribution. For example, a uniform distribution directly leads to a critical network. 
\begin{table}[h!]
    \centering
    \begin{tabular}{|c|c|}
        \hline
        Class of Function & Fraction ($\%$)  \\
        \hline
        Canalizing & 89.24 \\
        Nested Canalizing & 83.9 \\
        Symmetric &  98.86 \\
        2-Symmetric & 77.88 \\
        \hline
    \end{tabular}
    \caption{The fraction of features in biological networks disregarding functions with less than three inputs.}
    \label{tab:features}
\end{table}
\newpage

\begin{figure}
  \includegraphics[width=0.95\linewidth]{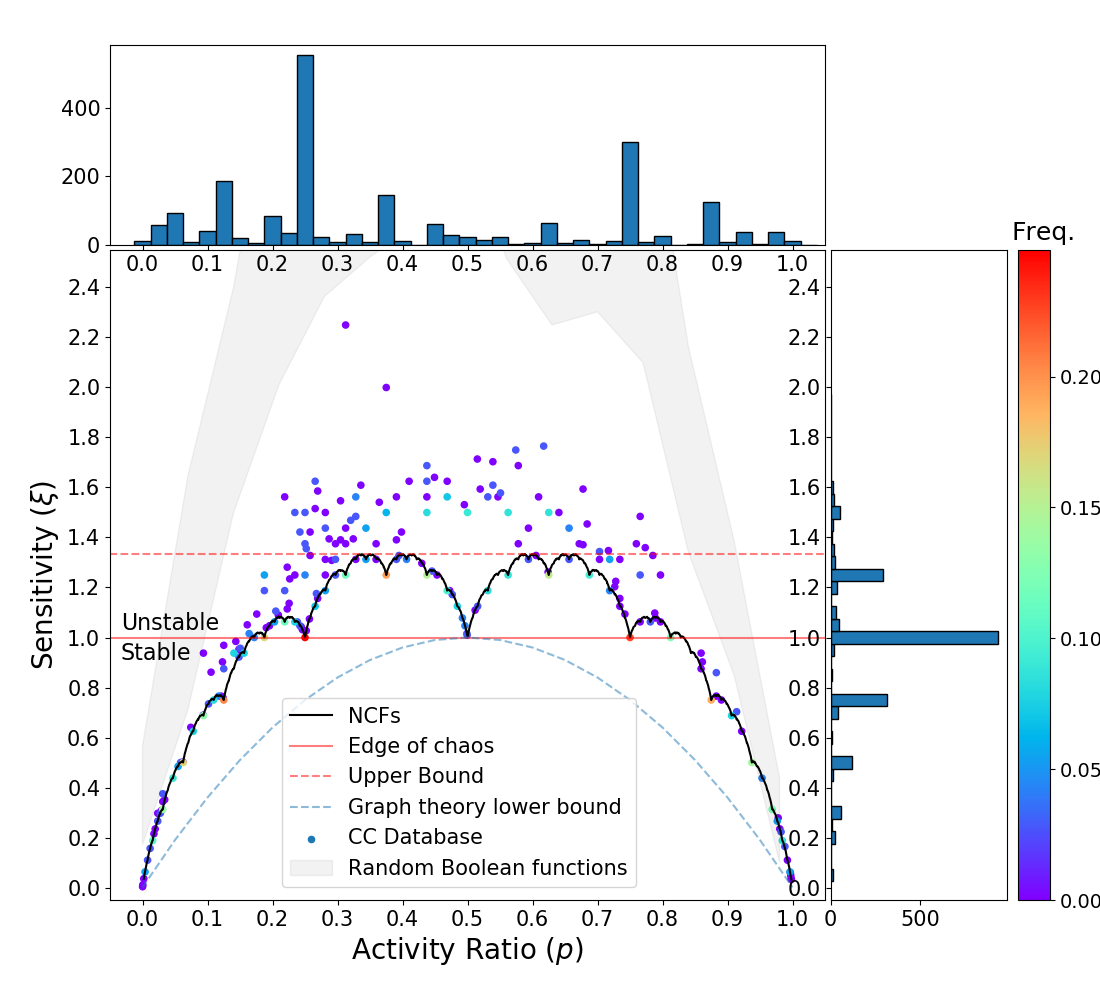}
  \caption{\label{fig_sensitivity} Sensitivity {\em vs} activity ratio
    for NCFs corresponding the minimum of $\xi(p)$ (solid) as proven
    here, and for biological examples ($1-$input functions are excluded) from Cell Collective
    database~\cite{helikar2012cell} (circle) with hot colors
    representing the higher frequency of occurrence in the
    database. Activity and sensitivity histograms for the latter are
    also shown on the side panels. Note that the sensitivity
    the histogram has been discussed earlier in
    Ref.~\cite{daniels2018criticality}. The shaded region corresponds
    to $1\sigma$ neighborhood of the mean sensitivity for randomized
    versions of the biological examples. The horizontal red line marks
    the stability boundary (edge of chaos). The lower bound adopted
    from spectral graph theory in the text is also shown (dashed). The upper bound for NCFs is the dashed red line \cite{li2013boolean}.}
\end{figure}
\newpage

\subsection{Sensitivity vs. Activity ratio}

The relation between sensitivity and activity ratio of special Boolean functions is discussed in the previous chapter, and here, this discussion is broadened with the numerical analysis of biological networks. In Fig. \ref{fig_sensitivity}, Boolean functions' sensitivities and activity ratios in biological networks and their frequencies are plotted together with the fractal NCF curve, sensitivities of random Boolean functions, and previously known bounds.
\begin{figure}[h!]
    \centering
    \includegraphics[width=
    \textwidth]{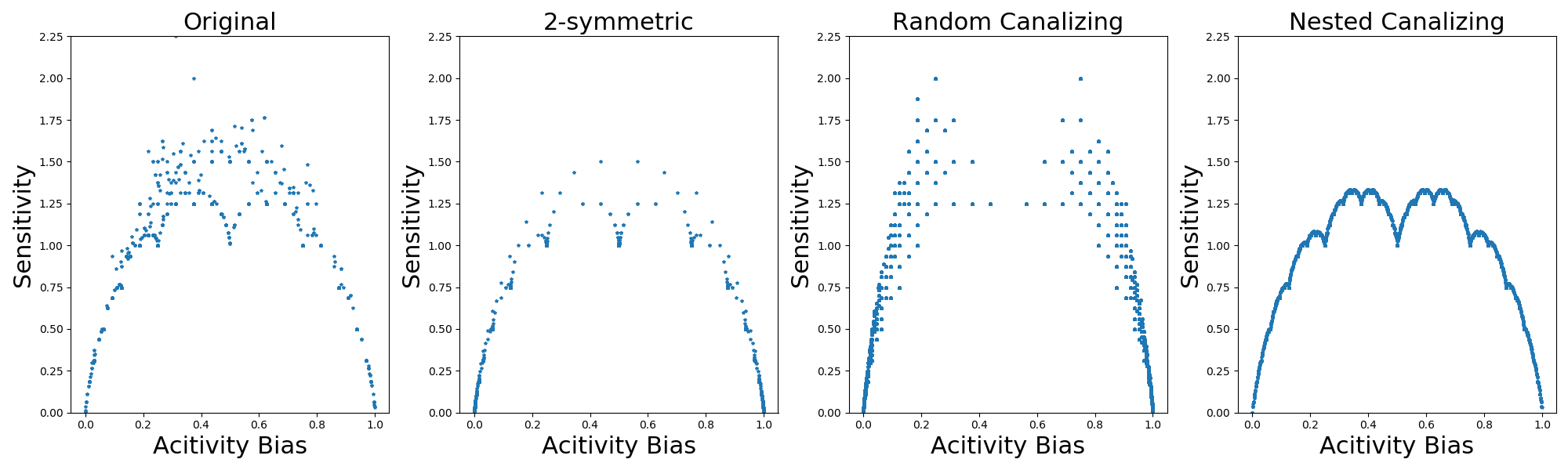}
    \caption{Sensitivities {\em vs.} activities for the original (Cell-Collective) gene regulatory functions and a random ensemble of canalizing functions. The full curves of {\em 2-symmetric} and nested canalizing functions are also generated.}
    \label{fig:sens_bias_features}
\end{figure}

The numerical analysis indicates that the biological functions have lower sensitivity than the randomly expected ones. 90 percent of the biological functions lie on the minimum sensitivity curve, and the non-NCF ones deviate from the minimum less than the random ones. Though the two third of the minimum curve stays in the unstable regime, the average sensitivity of the curve is equal to 1. The accumulation around the activity ratio of $1/4$ and $3/4$ is a consequence of  the abundance of $2-$input Boolean functions in biological networks. They also yield a peak in the sensitivity histogram.

Unlike canalizing functions, the expected sensitivity of \textit{2-symmetric} functions is independent of the number of inputs and depends only on the size of the small symmetry class, $a$. In the Cell Collective database, we only encountered $a=1,2,3$. For $a=1,3$ and $a=2$ the average sensitivities are $0.86$ and $0.946$, respectively, which fall in the vicinity of the critical regime. 
Interestingly, nested canalizing functions have an average sensitivity of exactly 1 under a uniform activity bias distribution. However, the corresponding fractal curve in Figure \ref{fig:sens_bias_features}, which is provably a lower bound~\cite{ccoban2022proof} on the sensitivity, does not capture a fraction of the data points from Cell Collective.
Random canalizing functions do not yield a better model either (also shown in Figure \ref{fig:sens_bias_features}) since the scatter above the lower bound is concentrated away from $\alpha=0.5$.
In contrast, \textit{2-symmetric} functions yield a sensitivity distribution that is in better agreement with the database.
They capture both the high density of genes on the boundary given by the nested canalizing functions and the scatter above the lower bound.

\begin{figure}[h!]
    \centering
    \begin{minipage}{0.5\textwidth}
        \centering
        \includegraphics[scale=0.4]{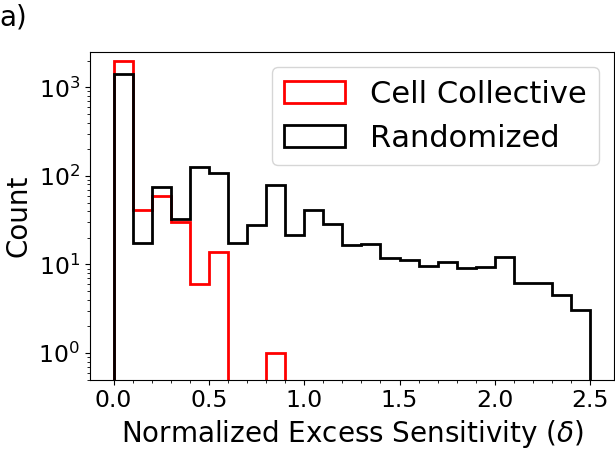}
    \end{minipage}\hfill
    \begin{minipage}{0.5\textwidth}
        \centering
        \includegraphics[scale=0.4]{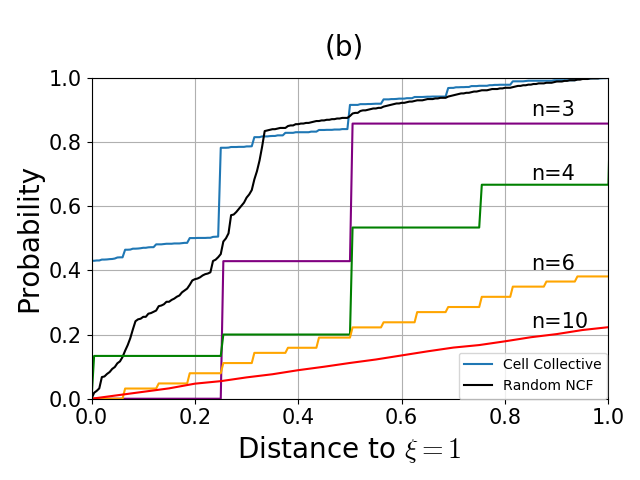}
    \end{minipage}
    \caption{\label{fig_histogram} (a) The histogram for the
      percentage deviation from the sensitivity minimum
      $\xi_{NCF}(p)$, for the regulatory functions in the Cell
      Collective database and their randomized counterparts. (b) The
      probability of finding $\xi \in [1-x,1+x]$ for functions in the
      Cell Collective database (blue), NCFs (black) and random Boolean
      functions with $n=3,4,6,10$ inputs.}
\end{figure}

Superimposing the resulting scatter plot with $\xi_{NCF}(p)$ in
Fig.\ref{fig_sensitivity} reveals the biological significance of the
calculated exact minimum. For comparison, the range of sensitivities
for an ensemble of randomized functions (obtained by shuffling the
truth table of each function in the database) is also shown as an
overhanging gray region in the same figure. It is remarkable that most
of the biological regulatory functions are situated on the minimal
curve and the rest are visibly closer to it than their random
counterparts. For a quantitative assessment of this observation, we
define the ``normalized excess sensitivity'' of a regulatory function
as $\delta[f]\equiv \xi[f]/\xi_{NCF}(p_f) - 1$.  In
Fig.\ref{fig_histogram}(a), we show the distribution of $\delta$ for the
functions in the Cell Collective database and for their randomized
versions. The dominating feature of the shown distributions is the
peak at $\delta=0$ which reflects the fact that, all but 215 functions
out of 2150 in Cell Collective lie on the sensitivity minimum (i.e.,
are NCFs, consistent with an earlier analysis on a smaller
set~\cite{kauffman2003random}). The remaining 10\% (non-NCFs) have
$\langle \delta\rangle \simeq 0.2$, as opposed to
$\langle \delta\rangle \simeq 0.85$ for the randomized functions.

\section{Conclusion}
It is interesting to consider our findings in conjunction with a
recent analysis on the same dataset by Daniels {\em et
  al.}~\cite{daniels2018criticality}, which found an impressive
concentration around the order-chaos boundary $\xi=1$ (also reproduced
here on side panel of Fig.\ref{fig_sensitivity}). The observation
serves as a confirmation of the well-known ``edge-of-chaos''
hypothesis by Kauffman, which posits that most biological systems
are tuned to remain in the vicinity of the critical
point~\cite{kauffman1969homeostasis,kauffman1993origins}, striking a
balance between robustness to transient environmental changes and
adaptability to persistent shifts. Mechanisms leading to criticality
in living organisms are still
unclear~\cite{vidiella2021engineering,MunozRMP2018}. Our results
underline the somewhat counterintuitive fact that, gene regulatory
networks not only ``live at the edge of chaos'', but they also barely
stray away from the boundary of minimum sensitivity. Upon inspection,
this is facilitated by the abundance of functions with two or three
inputs in the dataset (hot spots in Fig.\ref{fig_sensitivity}), which
induces a bias for certain activity ratios (Fig.\ref{fig_sensitivity},
top panel) with $\xi=1$. Yet, it is evident that the shape of
$\xi_{NCF}(p)$ favors the vicinity of the critical point, even in the
absence of any activity bias. Fig.\ref{fig_histogram}(b) shows that the
sensitivities of $50\%$ and $85\%$ of NCFs selected randomly from a
uniform distribution on $p$ remain within $1\pm 0.25$ and $1\pm 0.35$,
respectively.

We conclude that, a selective pressure for robustness combined with a
sufficient spread in the distribution of activity ratios suffices for
the gene regulatory functions to populate the neighborhood of marginal
stability, although an additional preference for a small number of
regulatory inputs per gene appears responsible for the sharp peak
observed at $\xi=1$ for biological
functions\cite{daniels2018criticality} (compare blue and black curves
in Fig.\ref{fig_histogram}). Thus, the exact bound $\xi_{NCF}(p)$ we
report here offers a quantitative reference point which helps one
gauge the role and the limits of the competition between robustness
and plasticity in shaping the marginal stability of these systems.

Finally, it is worth noting that, although the network sensitivity can
be expressed as $\langle \xi_\alpha \rangle$ (averaged over the
network nodes, $\alpha$) in an annealed approximation, existence of
correlations between the inputs of different nodes generally
necessitates a more refined
treatment~\cite{rohlf2002criticality,moreira2005canalizing,peixoto2010phase}. It
would be interesting to investigate the limits of sensitivity at the
network scale, in conjunction with the bound derived here at the node
level.

Biological benefits and origins of canalization in gene regulation have been discussed in the literature since the classic works of Waddington~\cite{waddington1942canalization} and Schmalhausen~\cite{schmalhausen1949factors}. Yet, the fractal lower bound on the sensitivity of Boolean functions (shown in the panel of Figure \ref{fig:sens_bias_features}), which is realized by nested canalizing functions, has been reported only recently~\cite{ccoban2022proof}. This observation poses the near-criticality of gene regulation dynamics as -almost- a mathematical necessity. On the other hand, a small but non-negligible portion of genes do not fall on this optimal boundary. While this can be interpreted as ``evolution in progress", an alternative perspective may follow from the analysis provided here. Biological networks appear to adhere more closely to generalized symmetry than to canalization in the regulation processes. Although these two concepts are not far from each other (as was shown in Figure  \ref{fig:venn}) asymmetric regulatory functions, unlike non-canalizing ones, are practically nonexistent.

We have shown here that this symmetry property alone yields an upper bound on the sensitivity (i.e., imposes robustness) of the regulatory dynamics, a desirable feature for any functional design. We then calculated the activity bias {\em vs.} sensitivity exhaustively for all {\em 1- and 2-symmetric} functions which correspond to the majority of the genes in Cell Collective. We finally argued that, a random ensemble of {\em 2-symmetric} functions may be a more faithful null-model for the biological networks than canalizing functions. Extending the above analysis to {\em p-symmetric} functions with $p>2$ and exploring biological mechanisms for symmetry in gene regulatory functions are left for future work.


\part{Gene Function Prediction using Gene Expression Profiles}

\section{Introduction and Literature Review}

Biology is a fertile ground for machine learning methods due to the huge demand for the automation of tasks such as data classification, anomaly detection, and image analysis, particularly in data-intense fields~\cite{ching2018opportunities}. Development of gene expression profiling techniques not only facilitated such automation in, for example, detection and classification of cancer from tissue samples, but also allowed identification of the relevant intra-cellular processes and genetic networks underlying a disease. A large input dimension and a relatively small number of samples are characteristic of such applications, which calls for appropriate tools such as support vector machines (SVM), Bayesian classification, random forest models, and, more recently, neural networks such as multilayer perceptrons (MLP), convolutional neural networks (CNN) or autoencoders (AE). The typical input for such models is either the microarray data (dominating the earlier literature) or the RNA-Seq profiles which are superior to microarrays in terms of sensitivity and specificity. With the volume of the RNA-Seq data growing fast, the more traditional supervised training approaches have started giving way to semi-supervised and unsupervised machine learning models (e.g., variational autoencoders) which have already shown promising performance on single-cell classification.

While predicting gene functionality and classifying of cells, gene expression profiles such as microarrays and RNA-Seq are the target for supervision in many statistical and machine-learning approaches. In the case of anomaly detection, there are several methods utilizing gene expression profiles to differentiate cancer cells. Ensemble supervised learning approaches with mainly support vector machines (SVMs) \cite{shipp2002diffuse} are shown to be effective by using the microarray gene expressions in this task \cite{tan2003ensemble}. A general review on exploiting microarray data is specified for statistical and machine learning approaches \cite{kuo2004primer}. Additionally, on the single-cell detection as a super-problem of cancer cell detection, well-known unsupervised and supervised machine learning methods like random forests, decision trees, K-means, and neural networks are compared and SVMs outperformed every other method \cite{pirooznia2008comparative}. Establishing the new sequencing technology "RNA-Seq" has partly obsoleted microarray gene expression profiles in terms of specificity and sensitivity. In general the effectiveness of unsupervised and supervised machine learning models has been discussed in a recent review \cite{jabeen2018machine}. The increasing number of data leads to the possibility to train deep learning models effectively. An unsupervised \cite{li2020deep} and a semi-supervised deep learning model \cite{xiao2018semi} has shown promising performance on single-cell classification task using RNA-Seq data. \\

 Another data-dependent task in the field of genomics is gene function prediction. There are both statistical approaches such as guilt-by-association method \cite{montojo2010genemania} and text-based frequency \cite{guan2018multi}, also machine learning ones such as SVMs, stacked auto-encoders and hierarchical models. Despite supervised machine learning models of single-cell classification tasks abundantly depend on gene expression profiles, in this task the prevalent supervision is the gene or associated protein annotations from various databases. One of the first handling to this task is supported by SVMs by using Gene Ontology (GO) annotations of genes and the scope of predictions is narrowed since the unannotated genes are discarded throughout the study \cite{vinayagam2004applying}. Combining labeled and unlabeled data in a semi-supervised model eliminates the narrow scope problem \cite{zhao2008gene}. 
in order to tackle various gene classification problems \cite{krishnakumar2022operonseqer}.
Gene expression profiles are a tool for some classification tasks of genes. A general classification of genes based on their contribution to diseases is tackled with multiple machine learning models \cite{asif2018identifying}. On detecting specific functionality of genes, a hierarchical deep learning classifier \cite{feng2018hierarchical} and a contrastive learning approach \cite{du2019gene2vec} are supervised by GO annotations. A recent study also incorporates the spatiality of genes into a hierarchical classifier \cite{pazos2022gene}. Though plentiful approaches using annotations and other features of genes, the impact of gene expression profiles is partly overlooked. In this regard, one of the first studies which utilize the microarray data to classify genes by their essentiality proposes an ensemble learning method \cite{plaimas2010identifying}. To detect cancer-related genes a similar approach is tested \cite{ghanat2017machine}. As a rare occasion, RNA-Seq profiles are exploited to identify operon pairs on the genome \cite{krishnakumar2022operonseqer}. However, predicting gene functionality incorporating gene expression profiles is still lacking.\\

We here propose a semi-supervised approach for gene function discovery by utilizing the Denoising Autoencoder (DAE) architecture. DAE was recently demonstrated to perform well for identifying functionally similar gene pairs in an unsupervised setting\cite{du2019gene2vec,doi:10.1128/mSystems.00025-15,TAN201763}. This approach has the advantage of using first-hand experimental data as the only input, however, for the very same reason, assigning function to genes based on their similarity to another requires a significant post-processing stage for integration with existing databases, resolving ambiguities, and the like. We show below that a semi-supervised approach can efficiently incorporate such tasks into the training of the network and allow a directed search for novel genes that take part in a particular cellular process. Below, we describe the model in detail and then use it to process the available transcriptome data for V. cholerae, E. coli and P. aeruginosa. We demonstrate that the method successfully recovers functional tags of randomly selected genes excluded from the training stage for testing purposes. Furthermore, we propose a function for several genes with no previously known role.

\section{Materials and Methods}

\subsection{Neural Network Model}

 Denoising Autoencoders (DAE) are simple architectures whose purpose is to discover a low-dimensional representation of the input. It does so by means of an encoder which maps the input into a "feature space" and a decoder which recovers the input from its features. The forward propagation stage of the DAE (from the input to the feature space) can be expressed as
\begin{equation}
    \vec{y}=\mathbf{D}\,(\mathbf{E}\,\vec{\delta}(\vec{x})+\vec{b}_E)+\vec{b}_D
\end{equation}
where $\vec{x}$ is the $n_g\times 1$ input vector (here, normalized RNA-Seq read counts for each gene), $\vec{y}$ is the output feature vector, $\vec{\delta}(\cdot)$ is an element-wise noise injector, $\mathbf{E}$ is the $n_h\times n_x$ encoder matrix, $\mathbf{D}$ is the $n_x \times n_h$ decoder matrix, and $\vec{b}_E$, $\vec{b}_D$ are the corresponding bias vectors. The input noise in this model is generated by the dropout procedure which replaces each element of the input vector $\vec{x}$ by zero with probability $p$.

The architecture can be used for both unsupervised and semi-supervised learning. The loss function for the unsupervised version is
\begin{equation}
    L(\mathbf{D},\vec{x},\vec{y})=\frac{1}{n_g}| \vec{y}-\vec{x} |^2+\frac{\lambda_1}{n_g n_h} \sum_{i,j} | D_{ij} |
    \label{unsupervised}
\end{equation}
where the last term corresponds to L1-regularization with larger values of the (positive) hyperparameter $\lambda_1$ enforcing sparser $D$ matrices. Note that, weight regularization on the decoder matrix is sufficient since noise injection by dropout simultaneously functions as a regularizer for the encoder matrix \cite{zhang2021understanding}. Alternative loss functions can be conceived (e.g., Tan {\em et al.}~\cite{TAN201763} use cross-entropy instead of mean squared loss for the first term and no regularization), however, we found that the loss given in Eq.(\ref{unsupervised}) yields the best performance in gene-pair association on our dataset (Supplementary Materials).

Unsupervised training on the transcriptome data has been observed to map the input vectors onto a "feature space", where some feature dimensions can be associated with certain biological functions. This allows one to, for example, deduce whether a pathogenic pathway has been activated~\cite{}. On the other hand, a global feature-to-function mapping after training the autoencoder requires significant amount of post processing, even under the (unjustified) assumption that a one-to-one association between feature dimensions and biological functions exists.

In order to acquire full control of the feature dimensions, we propose a semi-supervised DAE model by adding a supervised term to Eq.\ref{unsupervised}. The purpose of additional loss is to relate each feature to a known biological process. In other words, the existing gene annotation databases (KEGG or GO) are utilized to softly fix a subset of the entries in each decoder column to unity. For each column, these entries are chosen to be a random subset (and comprising a certain predetermined fraction) of the set of genes with a chosen functional annotation. Thereby, each feature dimension is assigned a function label {\em a priori}. This semi-supervised model is implemented by means of the following loss function:

\begin{equation}
    L(\mathbf{D},\vec{x},\vec{y})=\frac{1}{n_g}| \vec{y}-\vec{x} |^2+\frac{\lambda_1}{n_gn_h} \sum\limits_{i,j} | D_{ij} | |1-D_{ij}|+\frac{\lambda_2}{n_gn_h}\sum\limits_{i,j \in S_i} |1-D_{ij}|
    \label{supervised}
\end{equation}
The second term in Eq.(\ref{supervised}) now pushes the decoder towards a binary matrix, while the last term forces those entries of column $i$, which correspond to a selected subset ($S_i$) of genes towards unity. The sets $S_i$ are chosen randomly from the genes sharing a functional annotation in the chosen database. Preassigning existing annotations to the columns of the decoder matrix in this manner puts an additional constraint on feature selection, forcing a decomposition of the expression profiles into contributions from distinct (but possibly overlapping) sets of functionally linked genes. Genes sharing a common functional annotation are only partially available to the autoencoder, some genes are deliberately left out for testing and others waiting to be discovered.

\subsection{Data Acquisition and Processing}

The inputs to the DAE are vectors whose entries correspond to gene expression levels (read counts, $n_r$) obtained from RNA-Seq experiments. The model is tested on Escherichia coli K-12 (due to the high volume of available data, $n_r=3300$), Pseudomonas aeruginosa, and Vibrio cholerae N16961. Datasets were downloaded using the command-line tool {\em fasterq-dump} available in \href{https://github.com/ncbi/sra-tools/wiki/HowTo:-fasterq-dump}{GitHub}. The raw RNA-Seq data were then pre-processed and normalized by means of the software package {\em Salmon}~\cite{patro2017salmon}. 

\begin{figure}
    \centering
    \includegraphics[width=0.45\textwidth]{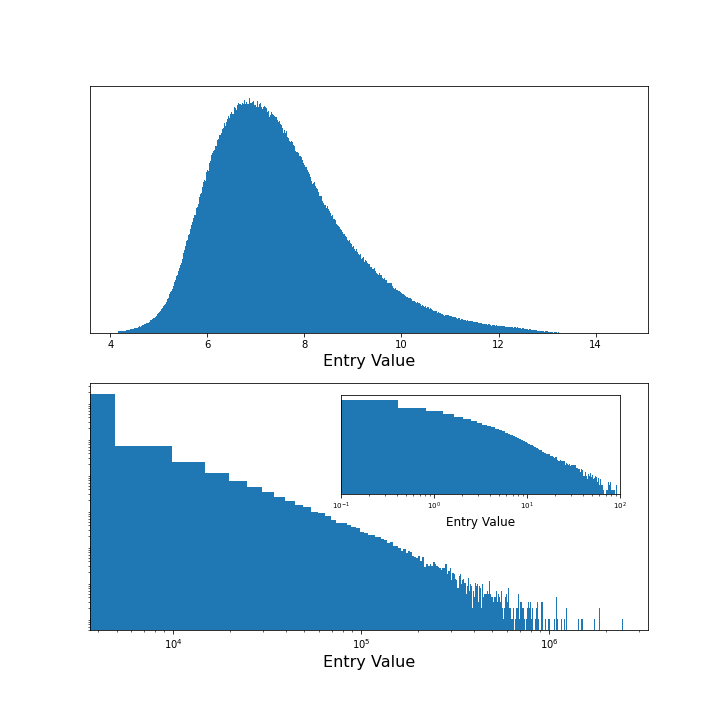}
    \caption{The histogram of entries in the data of P. Aureginosa(first figure) and VC (last figure) separately. The inset figure shows the histogram of normalized entries. The last figures are plotted in log scale. }
    \label{fig:data_shape}
\end{figure}

Mentioned strains of E. coli, P. aeruginosa, and V. Cholerae have $n_g = 4241, 5549, 3581$ genes, respectively (after discarding non-coding RNA sequences). Accordingly, the sizes $n_g\times n_r$ of the input matrices are $4241 \times 3300$ (E. coli), $5549 \times 950$ (P. aeruginosa) and $3581 \times 520$ (V. cholerae). It is performed a pre-normalization on the data to compensate for the fact that the mean read counts per gene span several orders of magnitude, as shown in Fig.\ref{fig:data_shape}. To this end, we rescale each row of the input matrix with its mean, so that the difference in several order of magnitudes between gene expressions. Such normalization is consistent with the first terms of Eqs.(\ref{unsupervised},\ref{supervised}) and helps the autoencoder avoid assigning undue importance to genes with high read counts and ignore those in the opposite extreme.

For model supervision and evaluation, we use either the KEGG pathway (\cite{10.1093/nar/gkv1070}) or the Gene Ontology (GO) annotation (\cite{ashburner2000gene}) databases. The information obtained for a particular organism from a database is represented by an $n_g \times n_a$ binary matrix, where each column corresponds to a pathway or a GO annotation and the nonzero entries in the column denote the genes that take part in the corresponding pathway or have the corresponding GO annotation.
The KEGG binary matrix, $K$, has the shape $4241 \times 124$ for E. coli, $5549 \times 169$ for P. aeruginosa and $3581 \times 111$ for V. cholerae, whereas the size of GO matrix for V. cholerae is $3581 \times 2421.$


\begin{figure*}[h!]
    \centering
    \includegraphics[width=\textwidth]{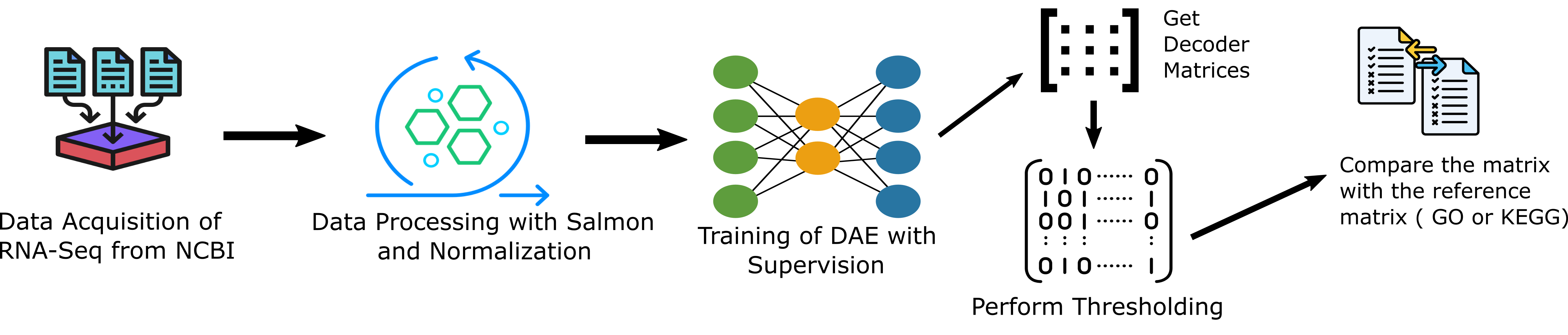}
    \caption{The pipeline of training and evaluation of a single DAE model.}
    \label{fig:pipeline}
\end{figure*}

\subsection{Evaluation Metrics}
To evaluate models there are two different metrics for unsupervised and semi-supervised approaches. For unsupervised case our metric is based on the method of Tan et al. (\cite{TAN201763}). Assigning each row of the decoder matrix to its corresponding gene we find $k$ closest gene for each gene. The distance between the row vectors calculated with cosine distance. The percentage of pairs which is included in at least one annotation is the success rate of our unsupervised models. This accuracy measure enforces to consider only annotated genes, since unannotated ones always fails to contribute to accuracy value. However we can include all genes to inspect possible relations including unannotated ones.

Unlike unsupervised models, semi-supervised ones are trained with ensembles and evaluated both collectively and individually. Each model is trained with a different 90 percent of the reference annotation matrix. At the end of the training, the decoder matrix $D$ is thresholded to filter out low correlations. Thresholding can be performed globally or for each column (pathway) or each row (gene). The recovery rate of the hidden part of the reference matrix (i.e. test set) by the thresholded decoder matrix is the performance measure of semi-supervised models. The 1s in the thresholded decoder matrix which do not appear in the reference matrix are the candidate relations, which may not be discovered yet, between a gene and an annotation. Each model produces a candidate list, and by training an ensemble of models we can make a frequency table of these candidates as a measure of confidence. The combined candidate list can be further filtered by removing rare candidates. In the end, the candidate list is evaluated qualitatively.
\begin{figure*}
    \centering
    \includegraphics[width=\textwidth]{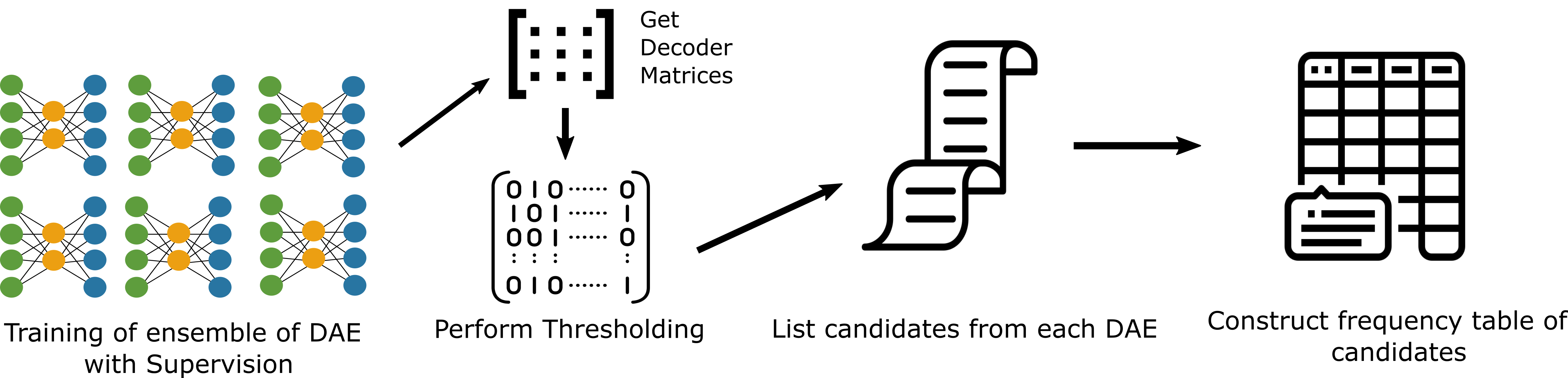}
    \caption{The pipeline of generating candidates from supervised models.}
    \label{fig:pipeline}
\end{figure*}

\section{Results}

\subsection{Structure of Models}

The data of VC consists of 520 vectors with a size of 3581. Previously by using microarray data of Pseudomonas Aureginosa unsupervised learning is performed with a similar DAE (\cite{doi:10.1128/mSystems.00025-15}). The distribution of entries in the microarray experiment data (with zero one normalization) has a positively skewed Gaussian distribution shape, while the distribution related to the VC RNA-Seq data indicates sparsity. Thus, the same models may not perform well simultaneously with these two different data. 

One modification in the proposed model is that the raw RNA-Seq data passed from Salmon Software should be normalized in such a way that the magnitude difference between the entries is minimized. Some vital genes usually have large entries due to their abundant activation, so we perform a gene-based normalization where each entry is divided by the mean activation of its corresponding gene. 

Another necessary component of the proposed DAE is the regularization of the weights. If the latent representation acquires dependencies of some biological processes or biological annotations i.e., the encoded vector captures the activities of biological processes, the decoder matrix should be sparse since the annotation matrices are also sparse. The optimal L1-regularization parameter in Eq.\ref{unsupervised} is found to be as $\lambda_1=0.623$. The regularization of the encoder matrix is done implicitly by applying a dropout layer before training as the noise component. 


The more complex the model becomes the higher number of parameters that should be adjusted correctly. Since our data is scarce the model should be as simple as possible, we stick to using only one hidden layer. Also, non-linearity is not introduced excluding the pre-normalization and the dropout layer. The models are trained with no activation function and mean squared loss.

\subsection{Ensemble of Semi-Supervised DAEs}

Randomness in our models, like random weight initiation and the dropout procedure, facilitates the advantage of an ensemble of DAEs (\cite{TAN201763}).

For VC and E.coli we train separate DAEs by using GO or KEGG annotations as supervision. Optimal hyper-parameter configuration in Eq.\ref{supervised} is found to be as $\lambda_1 = 0.74$ and $\lambda_2=0.01$. The latent dimension is adjusted according to the number of different annotations in the supervision set. While there are 111 different KEGG annotations for VC and 2421 GO annotations, 124 different KEGG and 4202 distinct GO annotations are present in the E.coli dataset. 

During the training randomly selected 90 percent of 1s of the annotation matrix is used as supervision; so, the latent dimension should be larger than the number of columns of the annotation matrix. For example, while some entries in the first 111 columns of the decoder matrix of a DAE trained with VC KEGG annotations are relevant for the supervised loss, the entries in other columns are not. Each column in the decoder matrix corresponds to a specific annotation. Thus, the latent representation should indicate the activation level of annotated biological processes. Choosing a latent dimension slightly larger than the number of annotations does not restrict models during the training, since there are also parameters that are not associated with any annotation.

\begin{figure}
    \centering
    \includegraphics[width=0.6\textwidth]{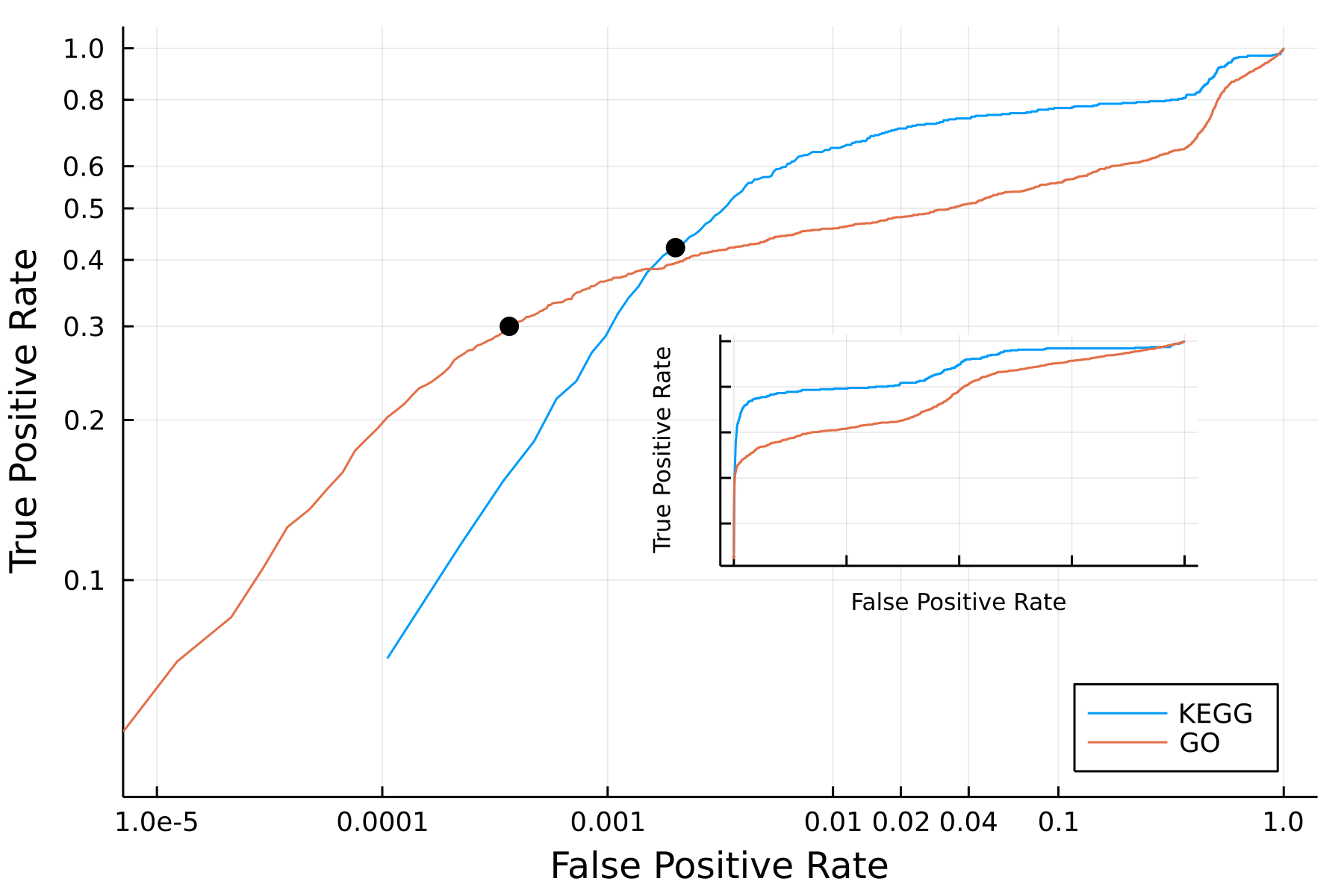}
    \caption{ROC curves for KEGG and GO-supervised ensembles with varying global threshold values. Log scale on the main figure, linear scale on the inset figure. The data points corresponding to selected threshold values are in black.}
    \label{fig:roc_supervised}
\end{figure}

The performance of the supervised ensembles is measured by their recovery rate which is mentioned in the chapter of evaluation metrics. Changing the threshold, the true positive and false positive rate of the annotating test set is depicted in Fig.\ref{fig:roc_supervised}. If conservative thresholds are chosen according to ROC curves such that 44.75\% additional KEGG annotations are predicted, the KEGG ensemble of 50 semi-supervised DAEs achieves an average recovery rate of 38,07 \%, which means that the models can predict 38,07 \% of the hidden 10 \% of the KEGG information on average while we enforce 44.75\% more annotations. The other 62\% should not be considered totally as false positives, they are rather predicted annotations for genes if they appear consistently in the ensemble.

The candidate list acquired from 50 different DAE can be sorted with their appearance rate as candidates. Also, we can filter the list so that only candidates related to the target KEGG annotations are left. If we use GO annotations instead of KEGG, the average recovery rate becomes on average 30.1 \% with the selected threshold.\\

To assess the predictive capability on unannotated genes, the ensemble with 10 models is trained without the annotated genes with single annotations, i.e. the single-annotated genes constitute to test set. The objective is to recover the hidden annotations using the candidate list approach. The predictions in the list are acquired by collecting the three highest weight-contributor pathways for each gene. There are 499 genes with one annotation in the KEGG database, while there are 3517 known annotations overall. For 49 genes, the most frequent predictions coincide with the correct one, and for 86 genes, the correct annotation is in the most frequent three predictions for these genes. Since from each model, three pathways are predicted for each gene, the probability of predicting the correct annotation of a gene by chance is $3/111\approx 2.7\%$. The number of correct predictions of a gene by a random ensemble is then expected to be distributed as a binomial distribution with probability $p=2.7\%$ and 10 trials. Moreover, for a given number of successes, the correctly predicted annotations have also a binomial distribution with 499 trials (genes). Comparing the performance of the proposed ensemble method with a random predictor, the annotations that are predicted by at least two models are statistically significant.
\begin{table}[h!]
    \centering
    \scalebox{0.77}{
    \begin{tabular}{|c|c|c|c|c|}
        \hline
      $\#$ of Success (Models)  &	Frequency &  Expected Freqeuncy &	Stdev of Frequency&	p-value\\
       \hline
        10&	7&	$10^{-13}$ &	$3.2\times10^{-7}$ &	0\\
        9&	12&	$3.7\times10^{-11}$&	$6.1\times10^{-6} $&	0\\
        8&	7&	$6\times10^{-9}$	&$7.75\times10^{-5}$&	0\\
        7&	6&	$5.77\times10^{-7}$&	$0.00076$&	0\\
        6&	9&	$3.64\times10^{-5}$&0.006&	0\\
        5&	2&	$0.00157$&	0.03967	&0\\
        4&	6&	$0.04725$&	0.217	&0\\
        3&	14&	$0.973$&	0.9855&	0\\
        2&	25&	$13.15$&	3.5783&	0.000464\\
        1&	69&	$105.312$&	9.1152&	0.999966\\
        0&	342&	$379.515$&	$9.5328$&	0.99996\\
    \hline
    \end{tabular}}
    \caption{\small The frequency of correctly annotated genes by a particular number of successes vs. random expectations and performed p-test of the observation on the ensemble method.}
    \label{tab:p_test_ensemble}
\end{table}

\subsection{Promising Candidates}

\subsubsection{Emsemble of Unsupervised DAEs}

The unsupervised approach can be used as a confidence source for the candidates of the supervised one. While applying the evaluation metric discussed in the previous chapter, the nearest neighbor genes should be found by using cosine distance. When weights corresponding to two almost identical genes are multiple of each other due to degrees of freedom in models, in terms of Euclidean distance they are far away from each other. However, in terms of their general behavior, they should be identified as similar, and capturing this missed association using cosine distance is crucial. 
Our model has similar accuracy values with the previously proposed model ADAGE.


Considering VC data, after a hyper-parameter search we find that the optimal choice for dropout probability is $p=0.01$ and the optimal hidden unit size is $h=200$. The best-performed model trained with MSE loss and feeding full batches of gene-based normalized data. The model includes L1-regularization ($\lambda_1=0.623$) and no activation functions. The nearest neighbor pairs are found with cosine distance and the accuracy of the best model is $0.6871$ on average.


While one single model produces approximately 1000 gene pairs, 67.9\% of them have a common annotation. If we train 50 DAEs and determine the frequencies of gene pairs, 431 gene pairs appear in every single model and 77.26 \% of them share a common annotation. One can generate more pairs by loosing the appearance rate condition of pairs with the cost of accuracy value. Including annotated and unannotated genes, three nearest neighbors of each gene are listed and they are utilized as a sanity check for our candidates from supervised ensembles.

\subsubsection{Novel Candidates}
The top ten predictions which persistently appear in the ensemble are listed in tables \ref{tab:kegg_cand} and \ref{tab:go_cand}. In table \ref{tab:go_cand}, the top ten predictions are aligned with the known role of the genes.  In tables \ref{tab:kegg_no_go} and \ref{tab:go_no_go}, the top ten predictions among the genes which have no GO annotations are listed as novel candidates. In this chapter, some notable candidates are presented.

Regarding one of our target annotations, Flagellar Assembly, the gene with locus tag VC1699 is a promising candidate, which appears 94.2 \% of the ensemble of DAEs trained with KEGG annotations. To increase the confidence of this prediction and to cross-check our models, we look at the 3 most common nearest neighbors of this gene using our unsupervised ensemble including also unannotated genes, since VC1699 is unannotated. Only 5 genes appear at least once in any model in the ensemble, the detailed information can be seen in the table.\ref{tab:1699_neighbors}. The gene with locus tag VC1022 appears in all models as a neighbor and it is related to the KEGG annotations Quorum sensing and Bio-film formation. The other neighbors are annotated with Flagellar assembly.  In 60\% of the supervised ensemble trained with GO annotations, at least one of the flagellum-related GO terms appears in the three highest weights corresponding to this gene.\\

\begin{table}[h!]
    \centering
 \begin{tabular}{|c|c|c|}
  \hline
      Neighbor& Frequency & Annotations  \\
      \hline
      VC1022&50& Quorum Sensing, Biofilm Formation \\
      VC2198&46& Flagellar Assembly \\
      VC2125&23& Flagellar Assembly \\
      VC2199&18& Flagellar Assembly \\
      VC2193&13& Flagellar Assembly \\
      \hline
 \end{tabular}
    \caption{The list of genes that is one of the three nearest neighbors of VC1699 in any model from the ensemble of 50 DAEs. Their frequencies among 50 models and annotations are also included.}
    \label{tab:1699_neighbors}
\end{table}

Another candidate VC2189 is again for Flagellar Assembly. This candidate appears in every model of the ensemble trained with KEGG annotations. By checking its three nearest neighbors in the unsupervised weights, only three genes appear in every single model: VC2199,VC2190 and VC2194. All three of them is annotated also with Flagellar Assembly. With 80\% probability, one of the three highest weight contributions among the weights of the gene from the supervised ensemble trained with GO annotation is related to the flagellum. \\

A novel relation for VCA1042 is also strongly suggested by our models. For all models in the KEGG ensemble, the pathway vch05110  Vibrio cholerae infection has the highest weight among the ones related to this gene. With 35\% probability, pathogenesis [GO:0009405] is one of the three highest weight contributors among the GO-Ensemble. The three nearest neighbors of the gene VC1043, detected by the unsupervised ensemble, are VC1456, VC0828, and VC0844 with 100\% confidence. VC1456 and VC0828 are linked with toxin proteins. VC0844 is poorly annotated and has no GO annotation. Although it is annotated with the KEGG pathway of biofilm formation.\\

VC1080 and VC2038, which produces phosphorelay protein LuxU, are predicted as related to phosphorelay sensor kinase activity [GO:0000155] and  phosphorelay response regulator activity [GO:0000156] by the ensemble trained with GO annotations. These candidates appear in at least 90\% of the models, respectively. The predicted annotations are significantly related to the true annotation of the phosphorelay signal transduction system [GO:0000160]. Another similar example is VCA0528, which is annotated as translation [GO:0006412]. The ensemble also suggests with 100\% confidence, that this gene is related to  translational elongation [GO:0006414] and translation elongation factor activity [GO:0003746]. These GO annotations are also related to the true annotation of the gene.\\

We can also focus on the non-annotated genes, to gain insight into their functionalities. Rather than thresholding cumulatively, thresholding for each gene separately guarantees at least one annotation for each gene. The most strict thresholding, which creates a one-hot vector, is applied for non-annotated genes to each decoder matrix. So, we only choose the annotation whose entry is the highest among others. Some non-annotated genes in GO database, VCA0118, VCA0119, VCA0121, and VCA0112 are predicted to be annotated as bacterial secretion system by almost all models in the ensemble. In fact, in the KEGG database they are annotated with this biological process. \\

Considering the KEGG ensemble, another set of non-annotated genes VC1748, VC1749, and VC1750 have the highest weights among their weights for the beta-Lactam resistance. GO-ensemble also suggests a relationship between these genes and the efflux pump. According to more than 85\% of the ensemble they are related to this pathway. Also with an unsupervised ensemble, we can ensure that their nearest neighbor genes are always VC1745, VC0165, and VC0166. VC0165 and VC0166 play a role in beta-Lactam resistance.

VC1395 and VC1396 are also interesting target genes to investigate deeply. The gene VC1395 contains an authentic frameshift according to the KEGG database. It is denoted as related to the protein CheY. Chemotaxis [GO:0006935]
is in the first three weight contributors for VC1395 in the 95\% of the GO ensemble, and the fraction is 85\% for VC1396. All nearest neighbors of VC1396 and VC1395 are related to bacterial chemotaxis [GO:0006935]. From the same region, VC1400 can be also related to chemotaxis.

Lastly, the region of VC0926 VC0928 VC0929 VC0930 VC0932 VC0933 VC0935 can be related to the polysaccharide biosynthetic process [GO:0000271] as suggested in Table \ref{tab:go_no_go}.

\section{Conclusion}

In this chapter, a novel ensemble-voting approach predicting gene functionality via denoising autoencoders is presented. The ensemble of semi-supervised models, which is trained with the RNA-Seq expression profiles and known annotations of genes, is able to recover some part of the functionalities in the test set, and it may reveal a part of the undiscovered functions. Some possible novel relations between biological processes and unannotated genes are proposed by the model, and waiting for the experimental verification.

\begin{table}[h!]
    \centering
    \scalebox{0.7}{
    \begin{tabular}{|c|c|c|c|c|}
    \hline
    Gene 	&  Pathway &	 Frequency&	 Main role &	 Subrole \\
    \hline
    \multirow{2}{*}{}VCA0909	&   Biosynthesis of siderophore  	&1.0	 &	Biosynthesis of cofactors,&	Heme, porphyrin, \\ &group nonribosomal peptides&&prosthetic groups, and carriers&and cobalamin\\\hline
    VC0841	&  Vibrio cholerae infection &	1.0	 & 	Cellular processes	& Pathogenesis\\\hline
    VC0840	&  Vibrio cholerae infection & 1.0	& 	Cellular processes & Pathogenesis\\\hline
    \multirow{2}{*}{}VC0776	&   Biosynthesis of siderophore  	&1.0	 &	Transport and binding proteins&	Cations \\ &group nonribosomal peptides&&&\\\hline
    VC2705 &  Propanoate metabolism &	1.0	 &	Transport and binding proteins &	Unknown substrate\\\hline
    VC2712 &  Purine metabolism	& 1.0 & 	Transport and binding proteins &	Nucleosides, purines and pyrimidines \\\hline
    VC0991 &  Nitrogen metabolism& 1.0 & 	Amino acid biosynthesis&	Aspartate family\\
    \hline
    \end{tabular}}
    \caption{Some top candidates by frequency of the ensemble of DAEs trained with KEGG annotations. }
    \label{tab:kegg_cand}
\end{table}

\begin{table}[h!]
    \centering
   \scalebox{0.7}{ 
    \begin{tabular}{|c|c|c|c|c|}
    \hline
    Gene 	& Predicted Annotation &	 Frequency& 	 Main role &	 Subrole \\
    \hline
    VC0837	& pathogenesis [GO:0009405] &	1.0 &	Cellular processes&	Pathogenesis\\\hline
    VC0843	& pathogenesis [GO:0009405] &	1.0 &	Cellular processes	&Pathogenesis\\\hline
    \multirow{2}{*}{}VC1054&	   3'-5' exonuclease activity  &	1.0	&	DNA metabolism& DNA replication\\ &[GO:0008408]&&& recombination, and repair\\\hline
    \multirow{2}{*}{}VC0108&	  DNA polymerase III complex &	1.0	&	DNA metabolism& DNA replication\\ &[GO:0009360]&&& recombination, and repair\\\hline
    \multirow{2}{*}{}VCA0044& aminopeptidase activity  & 1.0 & 	Protein fate&	Degradation of proteins,\\&   [GO:0004177]&&&peptides, and glycopeptides\\\hline
    \multirow{2}{*}{}VC2678& DNA repair [GO:0006281] & 1.0 & 	DNA metabolism&	DNA replication,\\&   transporter activity &&&recombination, and repair\\\hline
    \multirow{2}{*}{}VC0640	&   protein targeting [GO:0006605]	&1.0	&	Protein fate&	Protein and peptide \\ &&&&secretion and trafficking\\\hline
    \multirow{2}{*}{}VC2389&	 'de novo' pyrimidine nucleobase &1.0&	Purines, pyrimidines, &	Pyrimidine ribonucleotide \\ &biosynthetic process [GO:0006207]&&nucleosides, and nucleotides&biosynthesis\\\hline
    \multirow{2}{*}{}VC2043&	 chromosome [GO:0005694]	&1.0&	DNA metabolism&	DNA replication,\\&&&& recombination, and repair\\\hline
    VC2052&	 heme transport [GO:0015886]&	1.0&	Energy metabolism	&Electron transport	\\\hline
    \multirow{2}{*}{}VC1635&	 RNA modification [GO:0009451]&	1.0&	Protein synthesis&	tRNA and rRNA\\&&&& base modification	\\
    \hline
    \end{tabular}}
    \caption{Some top candidates by frequency of the ensemble of DAEs trained with GO annotations. }
    \label{tab:go_cand}
\end{table}

\begin{table*}
\parbox{.45\linewidth}{
    \scalebox{0.75}{
    \begin{tabular}{|c|c|c|}
        \hline
        Gene & Predicted Annotation & Freq.\\
        \hline
        VCA0119	 & Bacterial secretion system	&1.0\\
        VCA0118	 & Bacterial secretion system&	1.0\\
        VCA0112	 & Bacterial secretion system&	1.0\\
        VCA0110	 & Bacterial secretion system&	1.0\\
        VCA0106	 & Bacterial secretion system&	1.0\\
        VC2189	 & Flagellar assembly &	1.0\\
        VC1619	 & Styrene degradation &	1.0\\
        VC1572	 & Styrene degradation &	1.0\\
        VC1564	 & Vibrio cholerae infection &	1.0\\
        VC1417	 & Bacterial secretion system &	1.0\\
        VC1400	 & Styrene degradation &	1.0\\
        VC0933	 & Biofilm formation &	1.0\\
        VC0844	 & Vibrio cholerae infection &	1.0\\
        \multirow{2}{*}{}VC1572&	  Biosynthesis of siderophore &	1.0\\ &group nonribosomal peptides&\\
        \multirow{2}{*}{}VCA0976&	  Biosynthesis of siderophore &	1.0\\ &group nonribosomal peptides&\\
        \multirow{2}{*}{}VCA1065&	  Amino sugar and nucleotide &	1.0\\ &sugar metabolism&\\
        VC2207	&  Flagellar assembly &	0.9888\\
        VC2206	 & Flagellar assembly &	0.9888\\
        VCA0159	 & Tyrosine metabolism &	0.9775\\
        VC1699	 & Flagellar assembly &	0.9775\\
        VC1750	 & beta-Lactam resistance &	0.9438\\
        VC1748	 & beta-Lactam resistance &	0.9551\\
        VCA0689	&  Butanoate metabolism &	0.9663\\
        VCA0052	 & Ribosome	&0.9663\\
        VC2005	 & Flagellar assembly&	0.9663\\
        VC1807	 & Galactose metabolism&	0.9663\\
        VC1773	 & Histidine metabolism&	0.9663\\
        VC1272	 & Ribosome &	0.9663\\
        VCA0692	 & Biotin metabolism &	0.9551\\
        VCA0535	 & Nucleotide excision repair &	0.9551\\
        VCA0446	 & Arginine biosynthesis &	0.9551\\
        VCA0312	 & Nucleotide excision repair &	0.9551\\
        VCA0095	 & Ribosome &	0.9551\\
        VCA0095	 & Starch and sucrose metabolism &	0.9551\\
        \hline
    \end{tabular}}
    \caption{\small Some candidate genes which has no GO annotation acquired from the KEGG ensemble}
    \label{tab:kegg_no_go}}
    \qquad
    \parbox{.45\linewidth}{
    \scalebox{0.75}{
    \begin{tabular}{|c|c|c|}
        \hline
        Gene & Predicted Annotation & Freq.\\
        \hline
        VC0160	& negative regulation of protein transport& 	1.0\\
        VC0926	&polysaccharide biosynthetic process &	1.0\\
        VC0928	&polysaccharide biosynthetic process &	1.0\\
        VC0935	&polysaccharide biosynthetic process &	1.0\\
        VC1009	& vibriobactin biosynthetic process& 	1.0\\
        VC1506	&tryptophan biosynthetic process& 	1.0\\
        VC1506	 &fatty acid catabolic process& 	1.0\\
        \multirow{2}{*}{}VC1506&	 fatty acid beta-oxidation using &	1.0\\ & acyl-CoA dehydrogenase&\\
        VC1515	 &tetrahydrobiopterin biosynthetic process& 	1.0\\
        VC1592	 &arginine biosynthetic process& 1.0\\
        VC1642	 &histidine biosynthetic process &	1.0\\
        VC1997	&phospholipase activity &	1.0\\
        VC2552	 &histidine biosynthetic process &	1.0\\
        VCA0144	&tryptophan biosynthetic process &	1.0\\
        \multirow{2}{*}{}VCA0280&	  glycine decarboxylation via&	1.0\\ &glycine cleavage system &\\
        VCA0938	 &prosthetic group biosynthetic process &	1.0\\
        VC0789	 &citrate transport &	0.95\\
        VC0793	&citrate lyase complex &	0.95\\
        VC0926	&phage shock &	0.95\\
        VC0933	&polysaccharide biosynthetic process &	0.95\\
        \multirow{2}{*}{}VC1191&	 double-strand break repair&	0.95\\ & via homologous recombination &\\
        VC1191	 & SOS response &	0.95\\
        VCA0010 &	 response to cold &	0.95\\
        VCA0163	&tryptophanase activity &	0.95\\
        VCA0280	&glycine cleavage complex &	0.95\\
        VCA0738	& oligopeptide transport & 	0.95\\
        VCA0821	& 'de novo' protein folding &	0.95\\
        VCA0821	& unfolded protein binding &	0.95\\
        VCA0976	& vibriobactin biosynthetic process & 0.95\\
        VCA0976	& heme transport &	0.95\\
        VCA1050	& phenylalanine 4-monooxygenase activity & 	0.95\\
        VCA1073	&tryptophanase activity & 	0.95\\
        VC0047	&pilus assembly & 0.9\\
        VC0509	&metallopeptidase activity & 	0.9\\
        \hline
    \end{tabular}}
    \caption{\small Some candidate genes which has no GO annotation acquired from the GO ensemble}
    \label{tab:go_no_go}}
\end{table*}

\newpage

\part{Additional Discussion}
\section{Dynamics of Non-SGD Based Learning Algorithms}
\subsection{Introduction}

In the last few decades, machine learning has spread over a variety of applications and captured the attention of researchers from various disciplines. Deep learning models have reached beyond human skills in some games and tasks. One of the most popular approaches to training deep learning architectures is stochastic gradient descent (SGD). However, other non-SGD-based learning approaches have gained attention since some of them capture the properties of biological neurons more and at the same time perform reasonably against the mainstream SGD-based learning models. The dynamics of properties of such learning schemes have not been investigated in detail, unlike the SGD learning scheme. This chapter will discuss dynamics of non-SGD learning by also investigating a particular biologically plausible learning algorithm.

The generic structure of deep learning models consists of layers. In each layer, the input is multiplied with a matrix and it might pass through a nonlinear activation function afterwards. The output of $i^{th}$ layer is the input of $(i+1)^{th}$ layer and the next input is given with the equation,
\begin{equation}
    y^{(i+1)}=f\left(W^{(i)}y^{(i)}+b^{(i)}\right)
\end{equation}
The learning scheme updates the matrices $W^{(i)}$ and biases $b^{(i)}$ at each epoch by using the inputs and outputs,
\begin{equation}
    dW^{(i,t)}=W^{(i,t+1)}-W^{(i,t)}=\eta F(\{W^{(i,t)}\},\{b^{(i,t)}\},\mathcal{Y}) \ ,
\end{equation}
where $\mathcal{Y}$ represents the dataset. The SGD update is based on minimizing a predefined loss which is a function of weight matrices, biases, input, and output. However, we are particularly interested in the non-SGD, generic type of learning. 

\subsubsection{Langevin Equation and Fokker-Planck}

The performance of a deep learning model heavily depends on the size of the training dataset. Thus, passing the whole data through a model before each update epoch is computationally intractable. The common approach is splitting the dataset into batches $Y^{(t)}$ which is a representative subset of the whole data. Nevertheless, the random choice of batches creates a random noise on the update scheme. 
\begin{equation}
    dW^{(i,t)}=\eta F(\{W^{(i,t)}\},\{b^{(i,t)}\},Y^{(t)})+\xi^{(i,t)} \ .
    \label{langevin}
\end{equation}
The Eq.\ref{langevin} is the Langevin equation where a random force accompanies a known force. The stationary distribution of the weights depends on the distribution of the random noise and the time dynamics of the distribution are governed by the Fokker-Planck equation. By discarding the index of weights and the explicit dependence of weights on each other,
\begin{equation}
    \frac{\partial P(W,t)}{\partial t}=\nabla\left( -\eta P(W,t)F(W(t),Y(t))+\nabla  D(W)P(W,t) \right) \ ,
\end{equation}
where $P(W,t)$ is the probability distribution of weights at time $t$ and $D(W)$ is the diffusion matrix generated by the random noise. The source stochasticity during the learning is the batching process. The selection of a batch can be performed either with replacement or without replacement of the previously chosen batch. Recently, the stationary distribution is investigated in the SGD case assuming two different batching approaches \cite{adhikari2023machine}. It is possible the extend this to non-SGD learning schemes. They pointed out that for SGD-based learning, which minimizes the loss $L(W)$, the stationary state distribution is a Gaussian distribution with variance $\Sigma_{\alpha\beta}$, and mean $W_0$. The variance satisfies the equation,
\begin{equation}
    H_0 \Sigma + \Sigma H_0 = 2\eta^{-1}D_0
    \label{einstein_relation}
\end{equation}
where $H_{0,\alpha\beta}=\partial_\alpha \partial_\beta L(W_0)$ and $D_0=D(W_0)$.

\subsubsection{The Stationary Distribution of General Learning}
In the SGD case, the updating force is the gradient of a loss function, i.e. the gradient of a potential. If in a non-SGD case, the force can be written as a gradient of another function (force is conservative), the analysis of such a learning scheme is the same as the one of SGD learning. If the force is non-conservative, there is no potential associated with the force; however, a potential function may exist in a space where friction coefficients vary and depend on the spatial coordinates. Then the force can be written as
\begin{equation}
    \vec{F}(x)=\mu(x)\Vec{\nabla}U(x) \ ,
    \label{factorization}
\end{equation}
where $\mu(x)$ is space-dependent, positive-definite (no negative friction), symmetric mobility matrix, and $U(x)$ is the potential function. The analysis in \cite{adhikari2023machine} is also capable to present the stationary state distribution of such update schemes. The existence conditions of such factorization for any given force might be an open question, but even in the absence of such factorization, the only difference occurs in the Hessian matrix. The Eq.\ref{einstein_relation} becomes,
\begin{equation}
    H_0 \Sigma + \Sigma H_0^T= 2\eta^{-1}D_0 \ ,
\end{equation}
where $H_{0,\alpha\beta}=\partial_\alpha F_\beta (W_0)$ and $H_0$ is not necessarily symmetric.

\subsection{A Biologically Plausible Learning Algorithm}
In this section, an example of a non-SGD-based learning scheme, which results in a non-symmetric Hessian matrix, will be investigated. A modified version of Oja's update rule is proposed in the paper \cite{krotov2019unsupervised}.
\begin{equation}
    dw_{\alpha\beta}=g(h_\alpha)(v_\beta-w_{\alpha\beta}h_\alpha) \ ,
    \label{krotov_update}
\end{equation}
where $v_\beta$ is a component of the input vector $v$ and $h_\alpha=\sum_\beta w_{\alpha\beta} v_\beta$. The learning activation function $g(h)$ is designated such that it reinforces competition between hidden units. The learning scheme favors the most fired hidden unit and penalized the second most fired unit. In this manner, hidden units are forced to learn separate features. The function is defined as,
\begin{eqnarray}
    g(h_i)=
    \begin{cases}
        1 & i=K\\
        -\Delta & i=K-1\\
        0 & \text{otherwise} \ ,
    \end{cases}
\end{eqnarray}
where $h_K$ is defined as the strongest activated unit and $h_{K-1}$ is the second. 

It is possible to write the update force as depicted in Eq.\ref{factorization}. However, this factorization cannot be unique for arbitrary symmetric mobility matrix, but symmetric mobility restricts the choice of potential function $U$. The potential (loss) function must be invariant under the permutation of $\{w_{\alpha i},v_i\} \leftrightarrow \{w_{\alpha j},v_j\}$. The simplest choice for a nontrivial potential is 
\begin{equation}
    U=\sum_i h_i = \sum_i \sum_j w_{ij}v_j 
\end{equation}
Since the update in Eq.\ref{krotov_update} do not depend on other hidden units it is possible to solve $mu$ for only one hidden and then generalize to get the total structure of $mu$. If there is one hidden unit, then the potential reduces $=\sum_i w_iv_i$. It is also redundant to use learning activation function $g$. The update in Eq.\ref{krotov_update} reduces to
\begin{eqnarray}
    dw_i=v_i(1-w_1^2)-\sum\limits_{i\neq j}w_iw_jv_j \ .
\end{eqnarray}
Using Eq. \ref{factorization},
\begin{eqnarray}
    dw_i = \sum_j \mu_{ij} (\nabla U)_j =  \sum_j \mu_{ij} v_j
\end{eqnarray}
Combining both,
\begin{eqnarray}
    v_i(1-w_i^2)-\sum\limits_{i\neq j}w_iw_jv_j = \sum_j \mu_{ij} v_j
\end{eqnarray}
The mobility matrix defines the friction coefficients in the parameter space $\{w_i\}$, and it should depend only on the space parameters for appropriate physical discussion. Thus, the solution for the mobility matrix is then
\begin{eqnarray}
        \mu_{ij}=
        \begin{cases}
            1-w_i^2 & i=j\\
            -w_iw_j & i\neq j 
        \end{cases}
\end{eqnarray}
In the general scenario, there are multiple hidden units and there must be two indices for weights. Also, this requires four indices for the mobility matrix. To protect the matrix structure of the mobility matrix the weight matrix can be flattened, 
\begin{equation}
    W=
    \begin{pmatrix}
        w_{11}  \qquad & w_{12} & \dots & w_{1N} \qquad & w_{21} & \dots & w_{MN}
    \end{pmatrix}^T
\end{equation}
Then the mobility matrix $\mu$ is
\begin{align}
    \mu=
    \begin{pmatrix}
        A_1 & 0 & \dots & 0\\
        0 & A_2 & \dots &0   \\
        \vdots & \vdots & \ddots & \vdots\\
        0 & 0 &\ddots & A_M\\
    \end{pmatrix}_{MN \times MN}
\end{align}
and each block matrix $A_i$ is a square matrix with size $N\times N$:
\begin{align}
    (A_i)_{\alpha\beta}= 
    \begin{cases}
         g(h_i)(1-w_{i\beta}^2) & \alpha=\beta \\
         -g(h_i)w_{i\alpha}w_{i\beta} & \alpha\neq\beta
    \end{cases}
\end{align}

Spectral properties of mobility matrix carry insight on mobility modes and directional friction properties. The eigenvalues of a block diagonal matrix are the set of eigenvalues of block matrices. The eigenvalues of $A_i$ are the kernel of $A_i-\lambda I$. Assume $\lambda=1$ is an eigenvalue of $A_i$. 
\begin{align}
    A_1-I=
    \begin{pmatrix}
        -w_{1}^2 & -w_1w_2 & \dots & -w_1w_N\\
        -w_1w_2 & -w_{2}^2 & \dots & -w_2w_N  \\
        \vdots & \vdots & \ddots & \vdots\\
        -w_1w_N & -w_2w_N &\ddots & -w_N^2\\
    \end{pmatrix}
\end{align}
Notice that vector $(1,-w_1/w_2,0,\dots, 0)^T$ is in the kernel of $A_1-I$. Similarly $(1,0,-w_1/w_3,0,\dots, 0)^T$ is an eigenvector of $A_1$. Thus, $\lambda=1$ is a $(N-1)$-many degenerate eigenvalue. The last eigenvalue is $1-\sum_i w_i^2$ and corresponding eigenvector is $(w_1,w_2,\dots,w_N)^T$. Since this eigenvalue can be either positive or negative, the mobility matrix is an indefinite matrix. Mobility matrices are generally positive-definite since negative eigenvalues imply that the system moves in the opposite direction of the exerted force on it. Some Brownian motion and active matter systems are observed to exhibit negative mobility \cite{rizkallah2023absolute,eichhorn2002brownian}. 

The eigenvalue-eigenvector pair $\lambda=1-\sum_i w_i^2,\Vec{e}=(w_1,w_2,\dots,w_N)^T$ implies that if the weights are on the surface of $N-$dimensional sphere, the force on the direction $\Vec{e}$, which is perpendicular to the sphere, does not move the system. Assuming $w_i>0,v_i>0 \ \forall i$, $\Vec{e}\cdot \Vec{\nabla}L$ is always positive; thus, the force cannot be negative in the direction $\Vec{e}$. If the parameters are outside the sphere, the eigenvalue is negative and the parameters are pushed toward the sphere though the force is outwards the sphere. Similarly, if the parameters are inside the sphere, the eigenvalue is positive and again the parameters move toward the surface of the sphere. This analysis is valid for all matrices of $A_i$; thus, the learning converges to a point where $\sum_j w_{ij}^2 = 1, \ \forall i$.

\bibliographystyle{apalike}
\bibliography{MSc}

\appendix

\end{document}